\documentclass[10pt,a4paper,twocolumn, twoside]{article}
\RequirePackage{graphicx}
\RequirePackage[switch]{lineno}
\RequirePackage{color}
\RequirePackage[colorlinks,citecolor=blue,urlcolor=blue,linkcolor=blue]{hyperref}
\usepackage{graphics}
\usepackage{graphicx} 
\usepackage{dcolumn}
\usepackage{bm}
\usepackage{multirow}
\usepackage{longtable}
\usepackage{booktabs}
\usepackage[latin1]{inputenc}
\usepackage{caption}
\usepackage[affil-it]{authblk}
\usepackage{blindtext}
\usepackage{abstract}
\usepackage{amsmath}
\usepackage{footnote}
\usepackage{arydshln} 

\usepackage{textcomp} 
\usepackage{upgreek} 

\makeatletter
\renewcommand{\section}{\@startsection{section}{2}{0cm}{-\baselineskip}
{0,5\baselineskip}{\normalsize\bfseries}}
\renewcommand{\subsection}{\@startsection{subsection}{3}{0cm}{-\baselineskip}
{0,5\baselineskip}{\normalsize\slshape}}
\makeatother

\begin{document}

\title{Large-size sub-keV sensitive germanium detectors\\	 for the CONUS experiment\\ 
}

\author{H.~Bonet$\rm ^1$, A.~Bonhomme$\rm ^1$, C.~Buck$\rm ^1$, K.~F\"ulber$\rm ^2$, J.~Hakenm\"uller$\rm ^1$, G.~Heusser$\rm ^1$, T.~Hugle$\rm ^1$, J.B. Legras$\rm ^3$, M.~Lindner$ \rm ^1$,  W.~Maneschg$\rm ^1$, V. Marian$\rm ^3$, T.~Rink$\rm ^1$, T. Schr\"oder$\rm ^{4}$, H.~Strecker$\rm ^1$, R.~Wink$\rm ^2$
}

\date{\small \it 
$^1$Max-Planck-Institut f\"ur Kernphysik, Saupfercheckweg 1, 69117 Heidelberg, Germany \\
$^2$Preussen Elektra GmbH, Kernkraftwerk Brokdorf, Osterende, 25576 Brokdorf, Germany\\
$^3$Mirion Technologies (Canberra), 1 chemin de la Roseraie, 67380 Lingolsheim, France\\
$^4$on behalf of Preussen Elektra GmbH, Kernkraftwerk Brokdorf, Osterende, 25576 Brokdorf, Germany
\vspace{0.3cm}
\\ e-mail address:\\ {\tt conus.eb@mpi-hd.mpg.de}}
\vspace{0.3cm}

\twocolumn[
\begin{@twocolumnfalse}
\maketitle

\begin{abstract}
Intense fluxes of reactor antineutrinos offer a unique possibility to probe the fully coherent character of  elastic neutrino scattering off atomic nuclei. In this regard, detectors face the challenge to register tiny recoil energies of a few keV at the maximum. The \textsc{Conus} experiment was installed in 17.1\,m distance from the reactor core of the nuclear power plant in Brokdorf, Germany, and was designed to detect this neutrino interaction channel by using four 1\,kg-sized point contact germanium detectors with sub-keV energy thresholds. This report describes the unique specifications addressed to the design, the research and development, and the final production of these detectors. It demonstrates their excellent electronic performance obtained during commissioning under laboratory conditions as well as during the first two years of operation at the reactor site which started on April~1, 2018. It highlights the long-term stability of different detector parameters and the achieved background levels of the germanium detectors inside the \textsc{Conus} shield setup. 

\end{abstract}
\end{@twocolumnfalse}
]
\vspace{1.0cm}

\section{Introduction}
\label{chapter1}

Neutrinos are highly elusive particles. For low momentum transfer, however, neutrinos and their antiparticles can elastically scatter off atomic nuclei such that the outgoing nuclear wave functions add up coherently. This enhances the probability to detect neutrinos by three to four orders of magnitude compared to standard neutrino interaction channels such as elastic neutrino-electron scattering or the inverse beta decay. Even \linebreak though coherent elastic neutrino nucleus scattering \linebreak (CE$\nu$NS) was predicted in 1974 \cite{Freedman:1973}, it has eluded detection for four decades mainly due to one technological obstacle: the smallness of the nuclear recoil energy $E_{nr}$ (unit: eV$_{nr}$) released by the struck nucleus. Further, in certain detector types the collectible ionization energy $E_{ee}$ (unit: eV$_{ee}$) for signal processing can be quenched due to dissipation processes\textcolor{black}{, which are typically described by the Lindhard theory \cite{Lindhard:1963}}. A low detector energy threshold is therefore mandatory. Next to it, intense neutrino fluxes are still requested to benefit from a higher statistics. The most promising sources are pion decay at rest ($\pi$DAR) sources and nuclear reactors. The first type produces neutrinos ($\nu$) and antineutrinos ($\bar{\nu}$) of several tens of MeV and of different flavors, while the second type releases exclusively electron-$\bar{\nu}$s with energies below 10\,MeV. In this second case the recoil energies are in the keV$_{nr}$ region and more difficult to detect, but the expected neutrino flux at realistic di\-stan\-ces to a reactor core is typically higher. Furthermore, the coherency condition $\lambda \simeq R$ -- with $\lambda$ being the de Broglie wavelength of the neutrino, and $R$ the target's nuclear radius -- is better fulfilled.\\
\indent The first detection of CE$\nu$NS signals was achieved by the \textsc{Coherent} experiment using CsI[Na]- and Ar-based detectors \cite{coherent:2017,coherent:2021} at the Spallation Neutron Source (SNS), which is a $\pi$DAR source. A first CE$\nu$NS detection with reactor $\bar{\nu}$s is still pending. The \textsc{Conus} experiment aims at detecting CE$\nu$NS using reactor $\bar{\nu}$s at the Kernkraft\-werk Brokdorf (`KBR') \cite{kbr:contact}, in Brokdorf, Germany, which is a commercial nuclear power 
\begin{figure*}[h]
\begin{center}
  \includegraphics[width=0.80\textwidth]{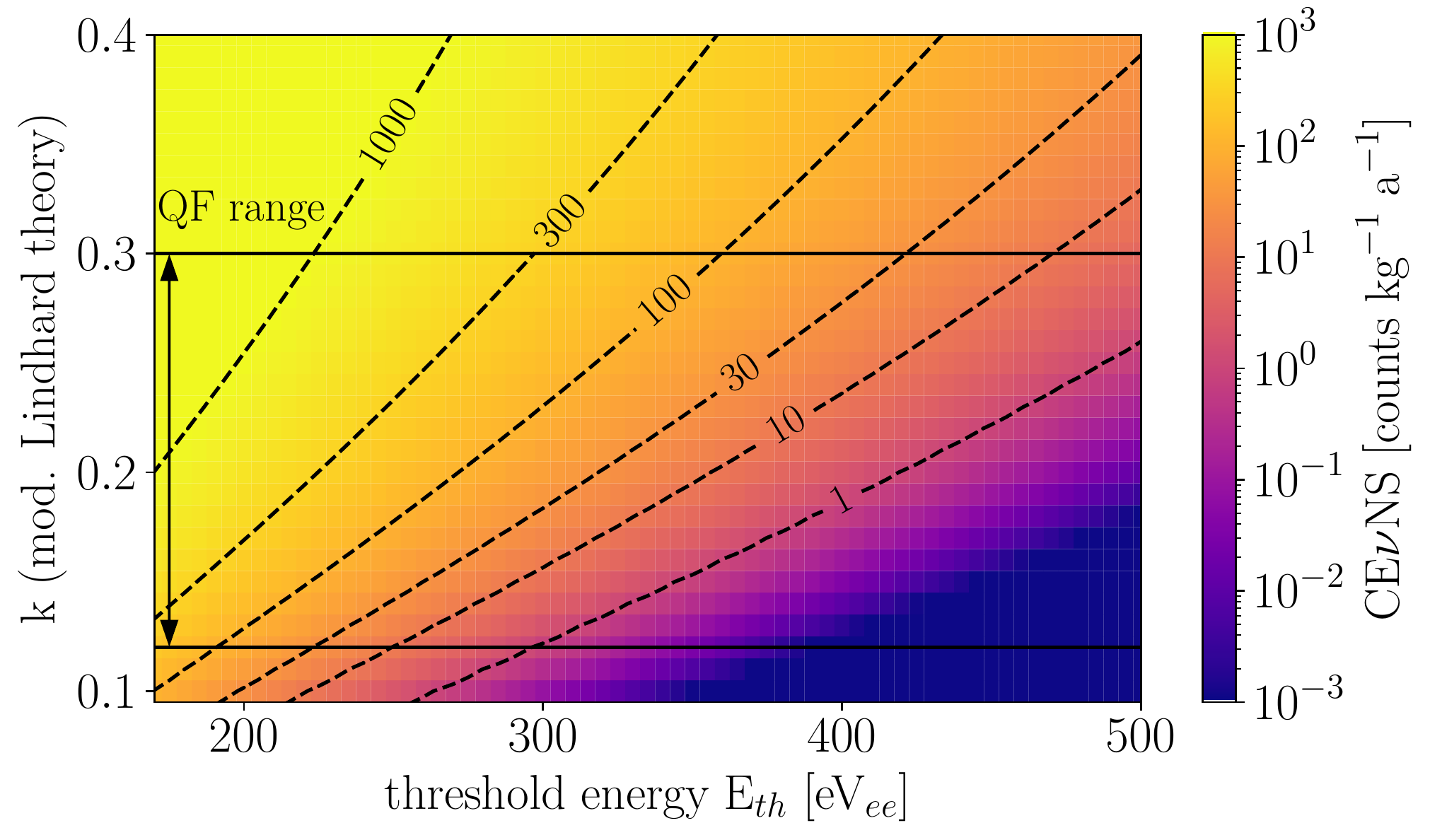}     
  \caption{\textcolor{black}{Expected CE$\nu$NS signal rate in \textsc{Conus} detectors at  17.1\,m distance from the reactor core with maximum thermal power $P_{th}$ of 3.9\,GW: the signal rate is depicted as function of i) the lower bound of the region of interest and thus the energy threshold, and ii) the experimentally not well known quenching factor (QF). Within the modified Lindhard theory, QF is described by the $k$ parameter. The black horizontal lines confine the range of measured QF values reported in literature.}}
  \label{fig:signalexpectation}  
  \end{center}
\end{figure*}
plant operated by the Preussen Elektra GmbH \cite{preussenelektra:contact}. The \textsc{Conus} setup is located at a distance of 17.1\,m from the reactor core \cite{Hakenmueller:2019}. With a maximum reactor thermal power $P_{th}$ of 3.9\,GW the expected $\bar{\nu}$-flux at the experimental site can be calculated according to Ref.~\cite{gemma:2007}, resulting in an integral flux of \textcolor{black}{2.3}$\times$10$^{13}$\,s$^{-1}$cm$^{-2}$.\\
\indent Prior to its installation at KBR in January 2018, the envisaged detector technology had to be prepared fulfilling several specific prerequisites. These include limitations due to reactor safety (robustness of setup, reduction of flammable materials, prohibition of multi-purpose usage of cryogenic liquids), stable and autono\-mous operation during data collection (no possibility of remote control, access only via multiple personal interlocks) and background suppression strategies (detector intrinsic background, $P_{th}$ correlated background, cosmic radiation at shallow depth, locally contaminated dust and airborne radon).\\
\indent Despite the challenge of coping with quenched nuclear recoils, we opted for the high purity germanium (HPGe) ionization detector technology, which has been well known at Max-Planck-Institut f{\"u}r Kernphysik, in  Heidelberg, Germany, (`MPIK') for many decades \cite{Heusser:1995wd,Heusser:2015}. Further we decided to optimize \cite{Salathe:2015} the p-type point contact (PPC) detector design, ori\-gi\-nally proposed by \cite{Luke:1985,Barbeau:2008,Barbeau:2009} and already deployed for keV$_{ee}$ and sub-keV$_{ee}$ physics applications by the CoGeNT \cite{Aalseth:2012}, \textsc{Cdex} \cite{Zhao:2013}, \textsc{Texono} \cite{Soma:2014}, \textsc{Majorana} \cite{Abgrall:2017} and $\nu$GeN \cite{Belov:2015} Col\-lab\-o\-ra\-tions. Together with the company Mirion Technologies (Canberra), in Lingolsheim, France, (`Mi\-rion-Lingolsheim') we were able to combine a unique set of detector specifications for the first time:
\begin{enumerate}
\item large crystal mass: 1\,kg,
\item excellent pulser resolution: $<$\,85\,eV$_{ee}$,
\item low energy threshold: $<$300\,\,eV$_{ee}$,
\item long cryostat arm: $>$\,40\,cm,
\item very low background components,
\item short cosmic activation time: $<$\,100\,d,
\item electrical cryocooler with noise-cancellation.
\end{enumerate}
\textcolor{black}{\indent To grasp the importance of points 1.-3., the expected CE$\nu$NS signal rate at the \textsc{Conus} location is shown in Figure~\ref{fig:signalexpectation} as function of two parameters: the lower bound of the region of interest (ROI) limited at lower energies by the detector energy threshold $E_{th}$, and the quenching factor (QF). The latter is expressed via the quenching parameter $k$ of the modified Lindhard theory with an adiabatic correction \cite{Barker:2013,Scholz:2016}. For a realistic (optimistic) $E_{th}$=300 (270)\,eV$_{ee}$ and a realistic $k$ value of 0.20, one expects 50 (100)\,counts kg$^{-1}$a$^{-1}$ reaching up to 500-600\,eV$_{ee}$. With a realistic  background of 10\,counts kg$^{-1}$d$^{-1}$ in the ROI resulting from points 4.-6. and from a shield similar to that in Ref.~\cite{Heusser:2015}, a signal-to-background ratio of 1:70 (1:35) is achievable. 
\begin{figure*}[h]
\begin{center}
  \includegraphics[width=0.80\textwidth]{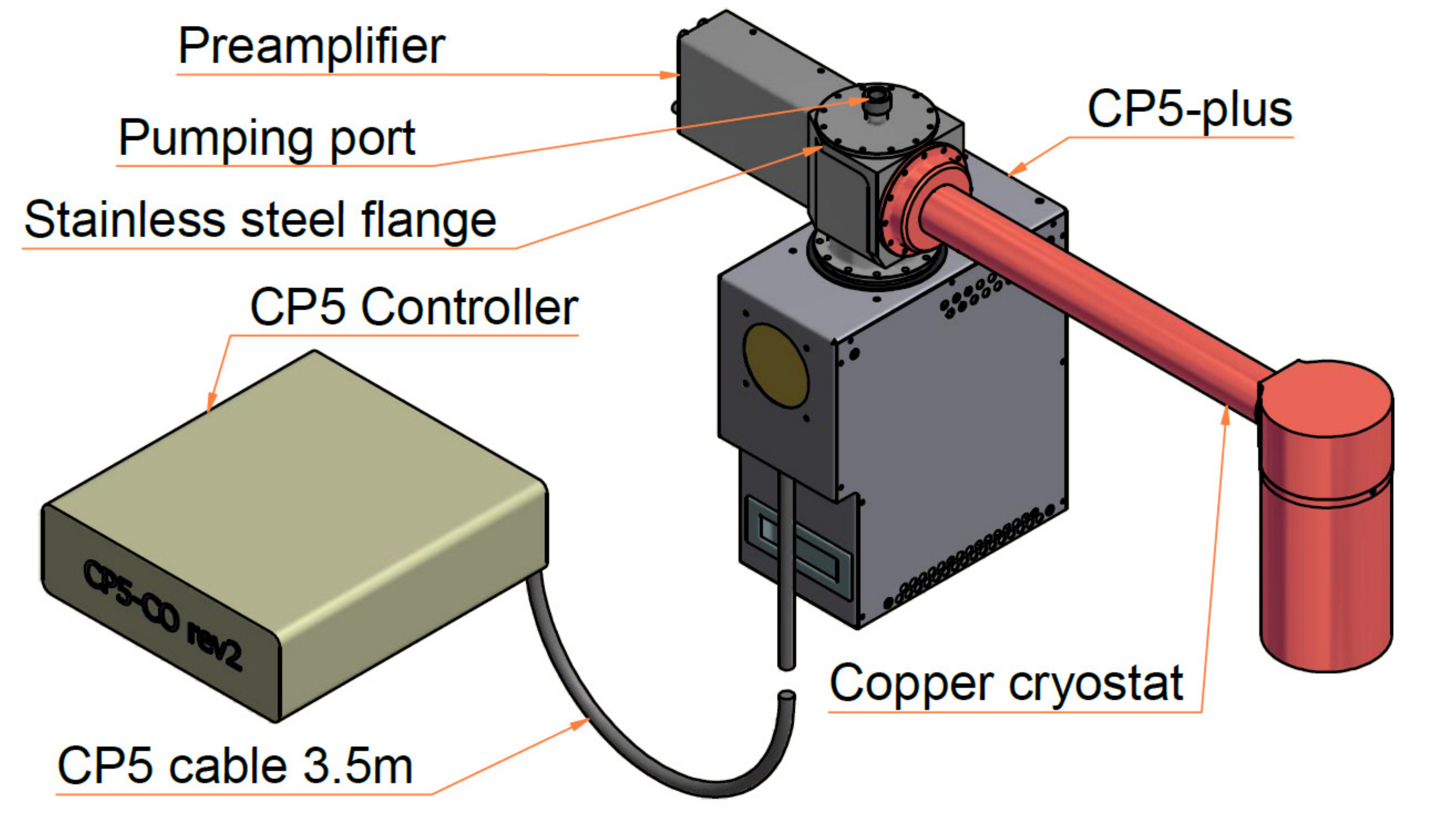}     
  \caption{Overall \textsc{Conus} detector design. The HPGe diode is included in the cylindrical copper end cap.}
  \label{fig:detectordesign_1}  
  \end{center}
\end{figure*}
By fulfilling these conditions CE$\nu$NS can become observable at the specified reactor site.}\\
\indent We pursued the construction of four PPC HPGe detectors, which were named \textsc{Conus}-1 to \textsc{Conus}-4 and are denoted with C1 to C4 in the following sections. Additionally a fifth PPC HPGe detector, C5, was built for research and development and for auxiliary measurements towards a better comprehension of the C1-C4 detector responses. C5 also has a crystal mass of 1\,kg and does not differ significantly in performance relative to the others, but it has been exposed longer to cosmic radiation overground. The following sections will focus only on C1 to C4, which underwent the same test procedures and were deployed simultaneously at KBR.\\
\indent This report is structured in the following way: Chapter~\ref{chapter2} illustrates the adopted overall detector design, i.e.\,~in cryostat, cryocooler and HPGe diode construction. It also includes the active volume, dead layer and transition layer characterization. Chapter~\ref{chapter3} gives an insight into the efforts to obtain intrinsically low background detectors for \textsc{Conus}. Chapter~\ref{chapter4} describes the electronics layout and the data collection system. Chapter~\ref{chapter5} focuses on the electronic response of the new detectors. This includes depletion voltage, energy scale calibration and linearity, energy resolution, trigger efficiency, as well as noise studies. Chapter~\ref{chapter6} refers to the long-term stability of the detectors evaluated especially during the experimental \textsc{Run-1} (April 1 -- October 29, 2018), \textsc{Run-2} (May 16 -- September 23, 2019) and partly \textsc{Run-3} (September 24, 2019 -- January 11, 2020) at KBR. Overall background rates will also be presented, however a detailed discussion of the background is postponed to a separate publication.\\
\indent To summarize, the excellent performance of the \textsc{Conus} detectors demonstrates the suitability of this detector design for CE$\nu$NS signal detection at reactor site.\\

\section{Detector design}
\label{chapter2}

In order to meet the low noise and low background prerequisites as well as the safety rules at the reactor site, the \textsc{Conus} Collaboration and Mirion-Lingolsheim cooperated strongly on the detector design and fabrication. The resulting detector design is shown in Figure~\ref{fig:detectordesign_1}. It can be divided into three main parts: the HPGe diode, the cryostat and the electrical cryocooler unit. Their specifications are discussed in the following subsections.

\subsection{HPGe diodes}
\label{High purity Ge Diodes}

\begin{figure}[h]
\begin{center}
  \includegraphics[width=0.40\textwidth]{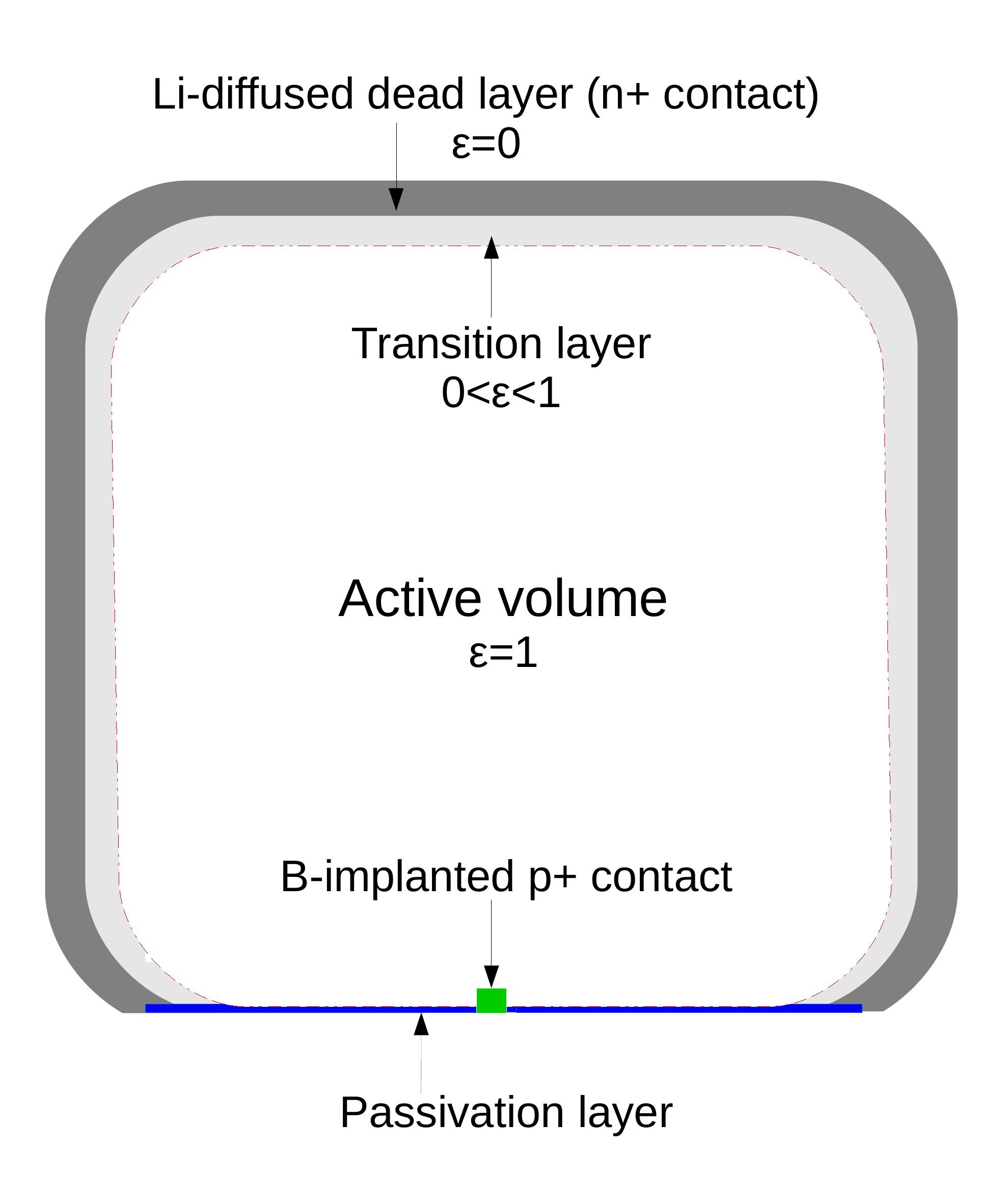}  
  \caption{PPC HPGe diode design adopted for the \textsc{CONUS} detector production. Both height and diameter are 62\,mm. Diode regions with different efficiencies $\varepsilon$ in signal detection are labeled accordingly.}
   \label{fig:diodedesign_1}  
  \end{center}
\end{figure}

\begin{figure}[h]
\begin{center}
  \includegraphics[width=0.50\textwidth]{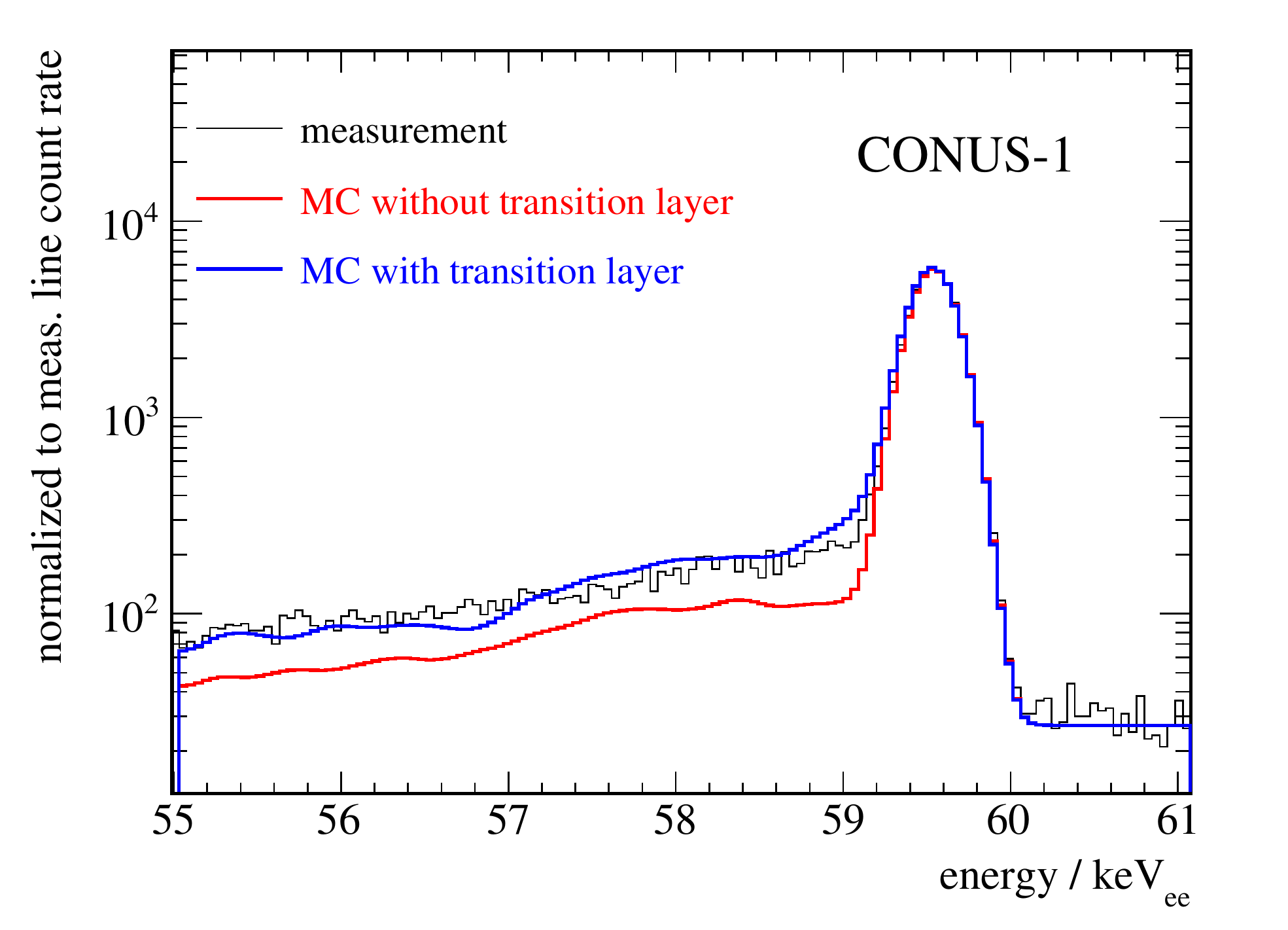}    
  \caption{Ionization spectrum of the C1 detector showing the 59.6\,keV$\gamma$-ray line from $^{241}$Am decays and the related continuum at lower energies. One MC simulated spectrum assuming a fully dead PCCL, and one distinguishing between a partially dead and a partially sensitive transition layer (TL) are shown. \textcolor{black}{Both are normalized to the measured line count rate, while the background is approximated with a flat component. The MC including the TL matches the experimental data better. A full agreement depends on a correct inclusion of all background contributions, especially of the external natural radiation and of cosmogenic interactions inside the used shield.}}
   \label{fig:TL_mc-vs-data}  
  \end{center}
\end{figure}

For HPGe diode production, five cylindrical p-type \linebreak HPGe crystals were ordered. All have a height ($h$) and a diameter ($\O$) of 62\,mm$\times$62\,mm, each corresponding to a total mass of 996\,g. In terms of noise suppression, the crystals and the net impurity concentrations were carefully selected in order to \textcolor{black}{reduce the bulk leakage currents} down to the sub-pA level.\\
\indent The HPGe diodes have a lithium-diffused n+ layer \linebreak wrapped around the lateral surface area and the top end area. As shown in Figure \ref{fig:diodedesign_1}, this layer is partially dead (dead layer (DL)) and semi-active (transition layer (TL)). Both layers together are called partial charge collection layer (PCCL). The bottom end area is fully passivated (passivation layer (PL)) except for a concentric spot of a few millimeters in diameter, where the boron-implanted p+ contact is located. The smallness of this electrode reduces the detector capacitance below 1\,pF at full electrical depletion and helps to reduce the noise level. Finally, the active volume (AV) confined by the PL and the PCCL has a full charge collection efficiency $\varepsilon$.\\
\indent Mirion-Lingolsheim contractually agreed on a PCCL thickness of [0.5,~1.0]\,mm, i.e.\,~an AV of [92,~96]\%, and estimated a PL thickness of [100,~200]\,nm. \textcolor{black}{After detector delivery, the \textsc{Conus} Collaboration focused on the determination of the DL, TL and AV values motivated by the following reasons: first, the AV is needed for the CE$\nu$NS count rate determination; second, the TL is responsible for the formation of so-called `slow pulses'. These occur when energy depositions from external particles (e.g. electrons/positrons or low energy photons emitted in muon-induced electromagnetic cascades) with small absorption lengths are stopped within the TL. The released charges diffuse slowly and lossy from the TL into the AV. Such signals have large rise times and significantly contribute to the low energy background spectrum}; third, the relatively wide PL is important for the background model of e.g.\,~surface $\alpha$-decays. These investigations were performed by using the measured background and radioactive source calibration data, and by comparing them with Monte Carlo (MC) simulations of the same detector setups. The \textsc{Geant4}-based simulation package \textsc{Mage} \cite{Boswell:2011} -- often applied and validated at MPIK in other occasions \cite{Hakenmueller:2019,Heusser:2015,Budjas:2009,Hakenmueller:2016,Agostini:2019} -- was used here.\\
\indent The PCCL thicknesses were measured with surface sensitive $^{241}$Am sources. The corresponding method is \linebreak based on a comparison of measured to simulated ratios of count rates in prominent $\gamma$-ray lines at 59.6, 99.0, and 103.0\,keV. For a comprehensive description of the applied method see Refs.~\cite{Agostini:2019,Agostini:2015,Lehnert:2016}. By modeling the TL with a proper sigmoidal-shaped analytical function in the MC, the observed low energy tail of 59.6\,keV \textcolor{black}{is better matched}; see Figure \ref{fig:TL_mc-vs-data}. The AV values were deduced from the known crystal volumes and the determined PCCL thicknesses. The overall results are summarized in Table~\ref{table:pccl_dl_tl_av}.\\

\begin{table}
\begin{center}
\caption{\rm{Active masses (AM), active volumes (AV) and thicknesses of the partial charge collection, dead and transition layers (PCCL, DL, TL) of C1 to C4.}}
	\vspace{0.3 cm}
	\begin{tabular}{lccccc}
	\hline
	\hline
	det.	   	&AM		  		&AV 		&PCCL 				&DL        	&TL 	\\
			  	&[g]	      	&[\%]		&[mm]				&[mm]      	&[mm] 	\\
	\hline
	C1			&936$\pm$10    &94$\pm$1	&0.77$\pm$0.03		&0.57		&0.20	\\
	C2			&947$\pm$10    &95$\pm$1	&0.69$\pm$0.03		&0.52		&0.17	\\	
	C3 			&936$\pm$10    &94$\pm$1	&0.64$\pm$0.03		&0.48		&0.16	\\
	C4			&907$\pm$10    &91$\pm$1	&1.32$\pm$0.04		&1.08		&0.24	\\
	\hline
	\hline
    \end{tabular}
	\label{table:pccl_dl_tl_av}	
\end{center}
\end{table}

\subsection{Cryocooler}
\label{Cryocooler}

To cool the HPGe diodes, we replaced standard liquid nitrogen cooling in favor of electrically powered pulse tube coolers. We selected the {\it Cryo-Pulse 5 Plus} model denoted with CP5+, which is offered as a second generation state-of-the-art product by Mirion Technologies \cite{mirion_cp5:contact}. \textcolor{black}{It is a Stirling pulse tube refrigerator \cite{DeWaele:2011}}, which was developed originally by Mirion-Lingolsheim and which turned out to widely fulfill reactor site limitations and several key specifications of a CE$\nu$NS sensitive detector.\\
\indent First, this cryocooler type is fully maintenance-free. It consists of a cold head assembly, to which the detector is directly attached, and uses a CFC-free, non-flammable gas in a hermetically closed containment, such that a gas refilling is not needed. The compressor contains no lubricant, which could pollute the refrigerator. Moreover, it has a demonstrated very reliable mean-time-to-failure of 1.2$\times$10$^5$\,h, which corresponds to a lifetime of more than 12\,a of operation \cite{mirion_cp5:contact}. This is long enough for the planned \textsc{Conus} experimental phase from 2018 to 2022.\\
\indent Second, it has an external power controller with data logger functionality and an integrated high voltage (HV) inhibit circuit.\\
\indent Third, the improved heat sinking allows operation at room temperatures up to 40\,\textdegree{}C. The latter two points are important especially during reactor outages, in which room temperatures close to the reactor core can exceed 30\,\textdegree{}C (cf.~Section~\ref{CONUS measurement conditions}) and power/ventilation failures might occur.\\
\indent Fourth, the CP5+ model has shock mounts and an active vibration cancellation system. This consists of an active feedback loop which measures the vibration level via an accelerometer mounted on the compressor. Based on the outputs from the accelerometer, a controller finely tunes the phase between the pistons of the pulse tube to minimize the vibration and thus the noise levels observed in the detectors. The remaining mechanical noise is mostly stable over time; however, if the room temperature changes significantly, the cryocooler has to counter act to keep a constant diode temperature; this can have an impact on the noise behavior (cf.~Section~\ref{Detector noise stability}).\\
\indent Fifth, the electrical cryocooler allows to modify the diode temperature such that it can be operated in a range of [-170, -210]\,\textdegree{}C. This degree of freedom is used to minimize leakage currents and thus optimize the energy resolution, 
but can also be applied for the study of temperature-dependent material parameters in Ge (drift velocities, Fano factor, average electron-hole-pair creation energy etc.) and to \textcolor{black}{cure long-term cryostat vacuum instabilities (cf.~Section~\ref{Detector noise stability})}. The applied diode temperature, the cryocooler cold head temperature and the power consumption of C1-C4 -- as measured at the beginning of the experiment in April 1, 2018 -- are summarized exemplarily in Table~\ref{table:cryocooler_temp_power}.\\

\begin{table}
\begin{footnotesize}
\begin{center}
\caption{\rm{Main \textsc{CONUS} detector cryocooler parameters at the beginning of data collection at KBR on April 1, 2018.}}
	\vspace{0.3 cm}
	\begin{tabular}{lccc}
	\hline
	\hline
	detector		&HPGe diode				&cold head			&power       	\\
	           		&temperature			&temperature		&consumption 	\\	
		 			&[\textdegree{}C; K]	&[\textdegree{}C]	&[W] 	\\
	\hline
	C1				&-185; 88				&38.8				&82				\\
	C2				&-195; 78				&45.1				&162			\\	
	C3 				&-185; 88				&40.1				&109			\\
	C4				&-195; 78				&44.4				&147		    \\
	\hline
 	\hline
    \end{tabular}
	\label{table:cryocooler_temp_power}	
\end{center}
\end{footnotesize}
\end{table}

\begin{figure}[h]
\begin{center}
  \includegraphics[width=0.50\textwidth]{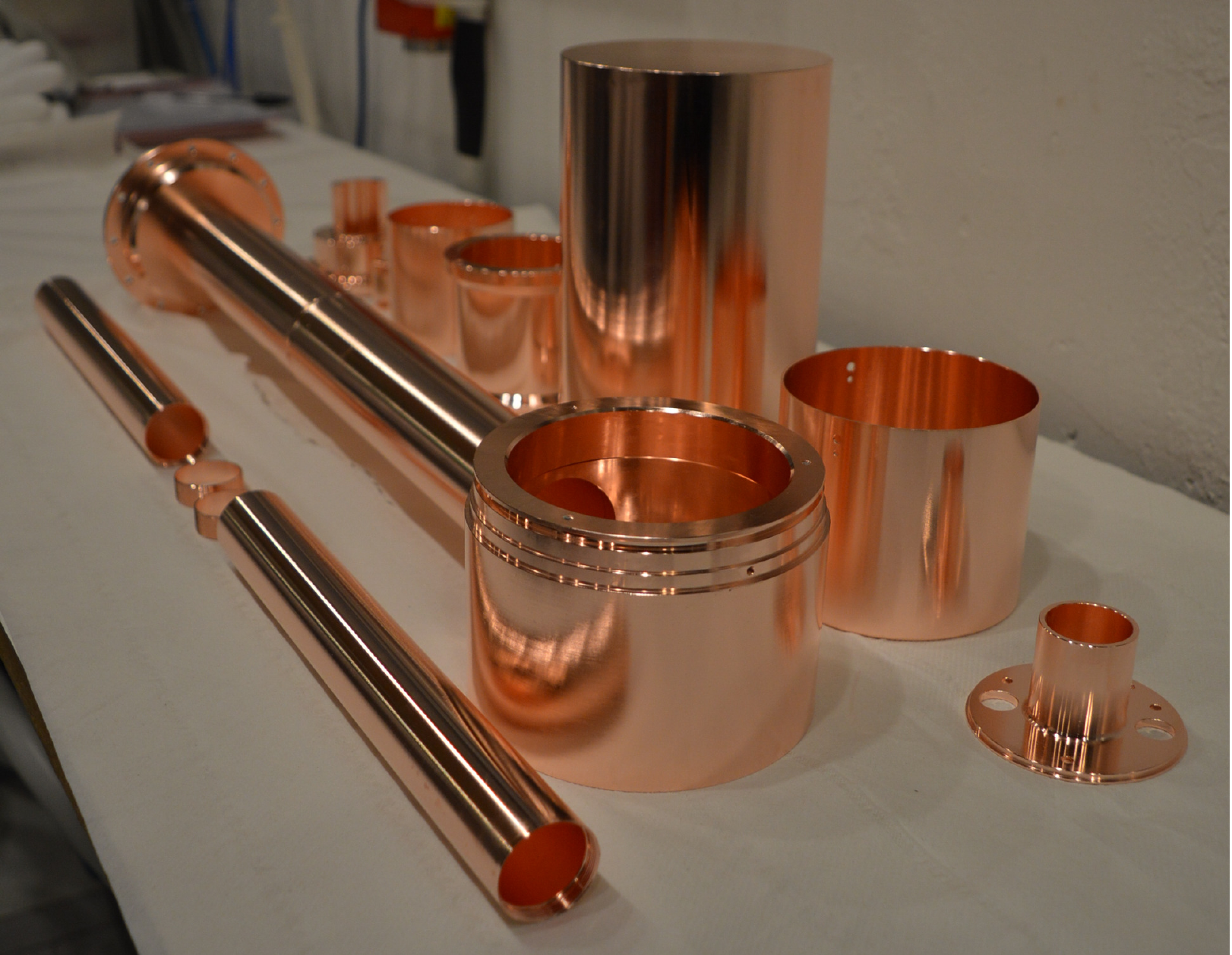}   
   \caption{Electropolished copper cryostat parts used for the construction of the first \textsc{Conus} detector C1.}
   \label{fig:cryostatparts_1}  
  \end{center}
\end{figure}

\subsection{Cryostat}
\label{Cryostat}

To shield the \textsc{Conus} HPGe diodes optimally  against external radiation of natural, man-made and cosmogenic origin with a proper shield setup (cf.~Section~4.1 in Ref.~\cite{Hakenmueller:2019}), a cryostat with an arm length of $>$\,40\,cm was required. For such lengths, electrical cryocoolers like CP5+ struggle with the coldness transfer, thus inducing more microphonic noise. In order to fulfill the strict noise, vacuum and background level specifications, Mirion-Lingolsheim spent a lot of effort in optimizing the cryostat design for \textsc{Conus}. This research was very successful, even though  some instabilities were encountered in a few detectors  during the long-term data collection period (cf.~Section~\ref{Detector noise stability}). The cryostats were made of electrolytic copper (Cu), wherein a few parts such as the end caps were refurbished from former low background HPGe detectors (cf. Sect. \ref{Lowbackgroundmaterials}). These end caps have an inner diameter of 105\,mm, which is more than sufficient to accommodate a 1\,kg massive HPGe diode and which has two further advantages: the large empty space allows for better cooling of the diode in all directions, and within a potential future upgrade, a HPGe diode of 1.5\,kg, i.e.\,~with dimensions 71\,mm$\times$71\,mm ($h\times\O$) could still be accommodated.\\ 
\indent The Cu cryostat parts were produced at the mechanical workshop of MPIK. Some manufactured Cu pieces are shown in Figure~\ref{fig:cryostatparts_1}. Finally, all Cu pieces were electropolished at the company \textsc{Poligrat}, in Munich, Germany \cite{poligrat:contact}, before being assembled, outgassed and commissioned at Mirion-Lingolsheim.

\section{Reduction of intrinsic detector background}
\label{chapter3}

\subsection{Search for low background materials}
\label{Lowbackgroundmaterials}

In terms of low background, the \textsc{Conus} experiment aims at a total background rate of $\leq$\,10\,counts kg$^{-1}$d$^{-1}$ in the interval [0.3,~1.0]\,keV$_{ee}$ including the ROI for CE$\nu$NS searches \textcolor{black}{(cf. Section~\ref{chapter1})}. This specification required not only a complex external shield design (cf. Section~4.1.2 in Ref.~\cite{Hakenmueller:2019}), but also a careful selection of all materials entering the detector production. This is discussed in the next paragraphs.\\
\indent First, the HPGe crystals were freshly grown (cf.~Section~\ref{High purity Ge Diodes}) and stored underground whenever not needed for HPGe diode processing. This helped to minimize the cosmogenic production of radioisotopes (cf.~Section~\ref{Cosmogenic activation in Ge and Cu}). Further, during HPGe diode production Mirion-Lingolsheim paid special attention not to accidentally introduce a contamination on the diode surfaces.\\
\begin{figure}[h]
\begin{center}
  \includegraphics[width=0.40\textwidth]{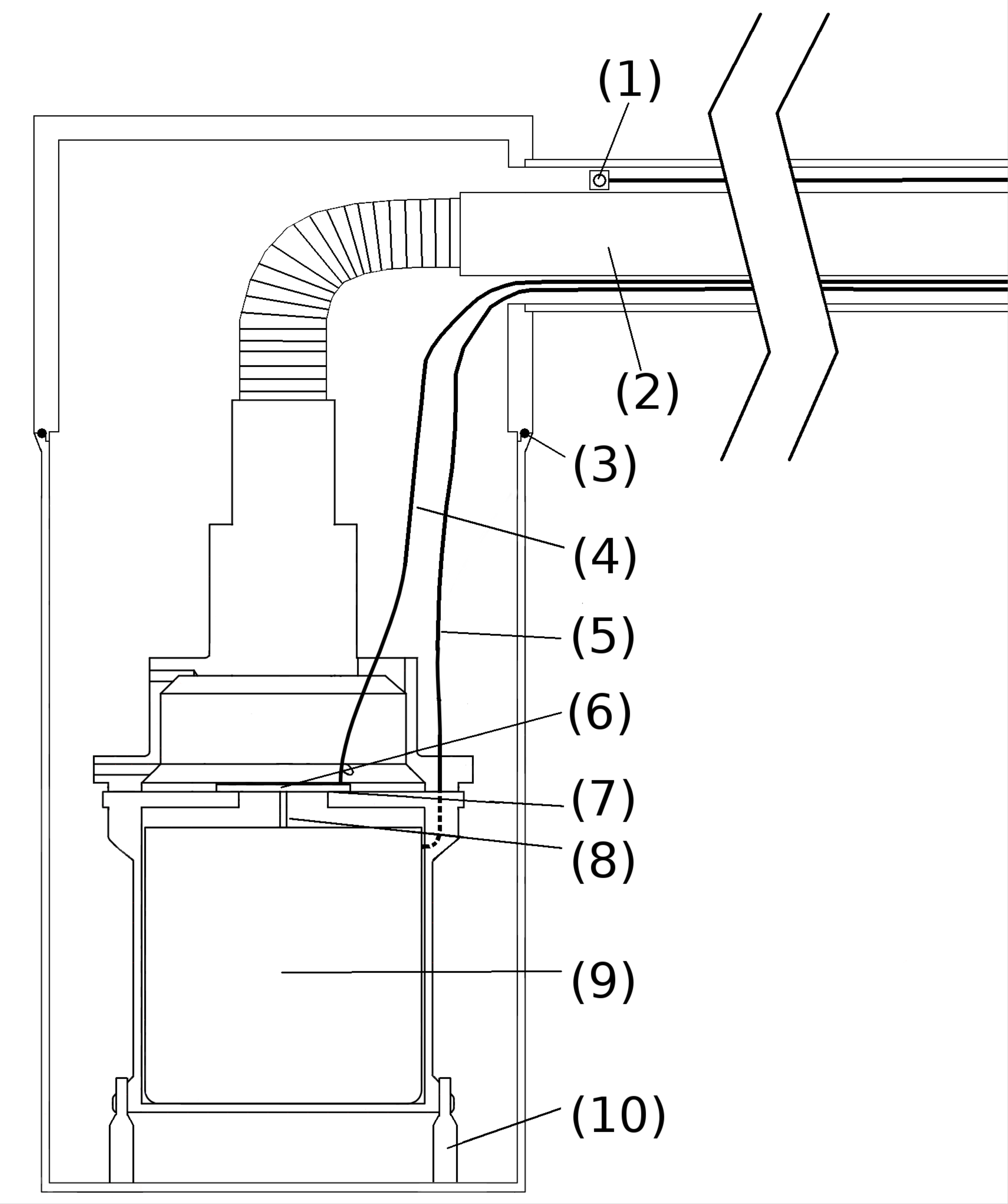}   
   \caption{\textcolor{black}{Illustration of the main components inside a \textsc{Conus} cryostat end cap: 1. temperature sensor, 2. cold finger, 3. O-ring, 4. signal cable, 5. HV cable, 6. substrate with cold front end electronics (SMD resistor, JFET with soldered contacts), 7. insulator, 8. contact pin, 9. HPGe diode, 10. PCTFE holder. All other components are made of electrolytic copper.}}
   \label{fig:endcap1}  
  \end{center}
\end{figure}
\indent Second, starting from samples put at the disposal by Mirion-Lingolsheim, the \textsc{Conus} Collaboration carefully measured the radio-impurity concentrations of all cryostat internal parts 
lying next to the HPGe diodes. \textcolor{black}{An illustration of these parts is given in Figure \ref{fig:endcap1}, while a list of the underlying materials together with their typical amount  integrated in a single detector are summarized in Table~\ref{tab:chap3_materialscreening}. The radio-impurity investigation} was performed by means of high sensitivity HPGe $\gamma$-ray spectroscopy, which has the advantage of being non-disruptive and enabling to collect information about progenies of the primordial U and Th decay chains as well. The latter point is important, since it proves the secular equilibrium among the sub-chains. This anticipates potential ratio changes that might occur within an experiment lasting several years such as \textsc{Conus}. Depending on the available mass and criticality, the samples were measured with the HPGe screening stations \textsc{Corrado} \cite{Budjas:2009} or \textsc{Giove} \cite{Heusser:2015} at the MPIK underground laboratory (15\,m of water equivalent, m w.e.),
\begin{table*}[hbt]
\begin{center}
\caption{Radio-impurity concentrations of the main material components used for the \textsc{Conus} detector construction. \textcolor{black}{Further, the typical quantities of materials deployed inside the end cap of a single detector unit are presented in the last column.}
\label{tab:chap3_materialscreening}}
\begin{tabular}{lccccccc}
\hline
\hline
sample        				&amount     &unit               & $^{226}$Ra & $^{228}$Th  & $^{228}$Ra    &$^{40}$K    &\textcolor{black}{amount/detector}\\
\hline
SMD resistor  				& 500\,pc.  & $\upmu$Bq/pc.     & 17$\pm$3   & $<$5        & $<$9          & $<$27      &\textcolor{black}{1\,pc.}\\
JFET (w/wo gold plating) 	& 100\,pc. 	& $\upmu$Bq/pc. 	& $<$0.4     & $<$2        & $<$2          & $<$10      &\textcolor{black}{1\,pc.}\\
PCTFE         				& 190\,g  	& mBq/kg            & $<$3       & $<$14       & $<$12         & $<$29      &\textcolor{black}{100\,g}\\
electronic substrate       	& 460\,g  	& mBq/kg       		& 13$\pm$2   & $<$5        & $<$4          & $<$17      &\textcolor{black}{10\,g}\\
HV cable      				& 6\,m     	& mBq/m        		& 0.7$\pm$0.4& $<$1        & 1.2$\pm$0.7   & $<$3       &\textcolor{black}{20\,cm}\\
signal cable  				& 70\,m    	& mBq/m             & $<$0.1     & $<$0.1      & $<$0.2        & $<$0.6     &\textcolor{black}{20\,cm}\\
soldering wire				& 104\,g   	& mBq/kg            & $<$42      & $<$27       & $<$45         & $<$124     &\textcolor{black}{$\sim$10\,mg}\\
O-ring        				& 26\,pc.   & $\upmu$Bq/pc.     & 570$\pm$120& $<$250      & $<$460        & $<$1500    &\textcolor{black}{1\,pc.}\\
insulator     				& 1.3\,kg   & mBq/kg  			& 4$\pm$2 	 & $<$2        & $<$4          & 40$\pm$12  &\textcolor{black}{5\,g}\\
contact pin   				& 100\,pc.  & $\upmu$Bq/pc.     & $<$5       & $<$8        & $<$9          & $<$19      &\textcolor{black}{1\,pc.}\\
\hline
\hline
\end{tabular}
\end{center}
\end{table*}
 as well as the \textsc{GeMPI} detectors \cite{Heusser:2006} at Laboratori Nazionali del Gran Sasso (LNGS, 3800\,m w.e.), in Assergi, Italy. They have increasing specific U and Th count rate sensitivities of 1\,mBq kg$^{-1}$, 100\,$\upmu$Bq kg$^{-1}$ and 10\,$\upmu$Bq kg$^{-1}$, respectively. The MC tool \textsc{Mage} (cf.~Section~\ref{High purity Ge Diodes}) was used to determine the sample-specific detector efficiencies and to predict the contribution of a given component to the total background in the final \textsc{Conus} setup. The main $\gamma$-ray screening results of accepted materials are reported in Table~\ref{tab:chap3_materialscreening}. Regarding the most challenging $^{228}$Th sub-chain (2614.5\,keV $\gamma$-ray emitter), only 
upper limits were found. In most cases, these are around or below 10\,mBq/kg. Regarding the U-related $^{226}$Ra sub-chain, only the electronic substrate has a finite concentration of (13$\pm$2)\,mBq kg$^{-1}$. In the case of the Viton${\textregistered}$ O-ring, which is needed to improve the long-term vacuum stability of the cryostat end cap, an additional $^{222}$Rn emanation measurement was performed at MPIK. The total rate for a sample consisting of 26 pieces was found to be acceptably low, i.e.\,~(757$\pm$42)\,$\upmu$Bq. Regarding $^{40}$K, only the insulator was found to have a higher concentration of (40$\pm$12)\,mBq kg$^{-1}$. In general, also $^{137}$Cs and $^{60}$Co concentrations were investigated, but are not reported here, since only small upper limits were found.\\
\indent Third, we used electrolytic oxygen-free high thermal conductivity (OFHC) Cu for the cryostat production. Contrarily to the default material aluminum, Cu is known to be highly radio-pure in the main primordial radioisotopes, but can be easily activated by the hadronic component of cosmic radiation overground. Thus, all Cu used for \textsc{Conus} was stored at MPIK underground laboratory whenever not needed for processing and assembling. Further, the ultra low background end caps from the Heidelberg-Moskow ex\-pe\-ri\-ment \cite{Guenther:1997} were refurbished, which have been stored at LNGS for two decades. A more detailed discussion of the Ge and Cu activation history follows in Section~\ref{Cosmogenic activation in Ge and Cu}.

\subsection{Cosmogenic activation in Ge and Cu}
\label{Cosmogenic activation in Ge and Cu}

\textcolor{black}{The dominant cosmic ray component at sea level is given by fast secondary nucleons with energies in the range of tens of MeV up to tens of GeV \cite{Hayakawa:1969,Ziegler:1996,Ziegler:1998}. Especially neutrons with such energies are able to induce spallation reactions in detector and shield materials, which produce long-lived radioisotopes \cite{Barabanov:2006,Elliot:2010}.} Thus, especially for components with larger masses and/or which are close to the detection medium, a stronger cosmic activation must be strictly avoided. To account for this, we used OFHC Cu either stored underground for decades or freshly produced. The electrolytic process applied for Cu can remove not only U and Th, but also cosmogenic radionuclides. Regarding the Ge material, we started from new Ge crystals. The segregation process \cite{Hall:1953} during Ge crystal pulling is very efficient in removing impurities including cosmogenic radioisotopes, which are not Ge isotopes like $^{60}$Co. In all cases we paid attention to keep materials overground only when needed for manufacturing. So, despite the complex manufacturing process the Ge was activated on average for 98\,d, and Cu for 102\,d. Finally we tracked the over-/underground storage periods of all Ge crystals / Cu parts until the final detector installation at KBR on January 24, 2018. With this information it is possible to estimate the expected activity of a radionuclide at a given time. The following two paragraphs focus on radioisotopes produced in Ge and Cu that can affect the Ge ionization energy spectrum below a few tens of keV$_{ee}$.\\
\indent In natural Ge, especially $^{68}$Ge, $^{65}$Zn and tritium ($^{3}$H) are relevant, but also $^{60}$Co, $^{58}$Co, $^{57}$Co, $^{55}$Fe, $^{54}$Mn and $^{51}$Cr were considered. Their half-lives and 
\begin{table*}[h]
\begin{center}
\begin{footnotesize}
\caption{\textcolor{black}{Activities $A$ (unit: decays kg$^{-1}$d$^{-1}$) of relevant cosmogenic radioisotopes in natural Ge included in the \textsc{Conus} HPGe diodes for the points in time T1=April 1, 2018 (beginning of \textsc{Run-1}) and T2=May 16, 2019 (beginning of \textsc{Run-2})}. The predictions are based on known activation and `cooling down' periods during detector manufacturing, on the reported half-lives and sea-level production rates $P$ (unit:  atoms kg$^{-1}$d$^{-1}$) that were averaged from \textcolor{black}{Refs. \cite{Ziegler:1998,Barabanov:2006} and Refs.~\cite{Avignone:1992}-\cite{Agnese:2019}}. The partly large standard deviations $\Delta P$ demonstrate the lack of precise measurements. 
\label{tab:chap3_cosmicactivation_ge}}
\begin{tabular}{lccccccccc}
\hline
\hline
isotope	&	$^{68}$Ge	&	$^{65}$Zn	&	$^{3}$H	&	$^{60}$Co	&	$^{58}$Co	&	$^{57}$Co	&	$^{55}$Fe	&	$^{51}$Cr	&	$^{54}$Mn	\\
\hline																			
half-life / d	&	270.0	&	244.3	&	4489.5	&	1934.5	&	70.9	&	271.8	&	985.5	&	27.7	&	312.0	\\
$P$	&	59.1	&	50.1	&	78.1	&	3.9	&	9.3	&	7.6	&	5.8	&	4.2	&	2.6	\\
$\Delta P$	&	43.9	&	21.9	&	57.4	&	1.5	&	4.5	&	3.9	&	2.6	&	--	&	1.7	\\
\hline																			
C1: T1	&	8.61	&	7.38	&	1.54	&	0.17	&	0.58	&	1.11	&	0.43	&	0.04	&	0.37	\\
\hspace{0.6 cm}T2	&	3.01	&	2.31	&	1.45	&	0.14	&	0.01	&	0.39	&	0.32	&	0.00	&	0.15	\\
\hdashline
C2: T1	&	6.32	&	5.44	&	1.13	&	0.12	&	0.61	&	0.81	&	0.31	&	0.09	&	0.27	\\
\hspace{0.6 cm}T2	&	2.21	&	1.70	&	1.06	&	0.10	&	0.01	&	0.29	&	0.24	&	0.00	&	0.11	\\
\hdashline
C3: T1	&	5.25	&	4.60	&	0.77	&	0.08	&	0.59	&	0.67	&	0.22	&	0.04	&	0.22	\\
\hspace{0.6 cm}T2	&	1.83	&	1.44	&	0.72	&	0.07	&	0.01	&	0.24	&	0.17	&	0.00	&	0.09	\\
\hdashline
C4: T1	&	6.56	&	5.71	&	1.05	&	0.11	&	0.67	&	0.84	&	0.30	&	0.04	&	0.28	\\
\hspace{0.6 cm}T2	&	2.29	&	1.78	&	0.98	&	0.10	&	0.01	&	0.30	&	0.22	&	0.00	&	0.11	\\
\hline
\hline
\end{tabular}
\end{footnotesize}
\end{center}
\end{table*}
\begin{table*}[h]
\begin{center}
\begin{footnotesize}
\caption{Activities $A$ (unit: decays kg$^{-1}$d$^{-1}$) of relevant cosmogenic radioisotopes in Cu used for the \textsc{Conus} Ge detector cryostats for the points in time T1 and T2. The prediction is based on known activation and `cooling down' periods during detector manufacturing, on the reported half-lives and sea-level production rates $P$ (unit:  atoms kg$^{-1}$d$^{-1}$) that were averaged from Ref.~\cite{Mei:2009} and Refs.~\cite{Back:2008}-\cite{Zhang:2016}. Several standard deviations $\Delta P$ are once again large.  
\label{tab:chap3_cosmicactivation_cu}}
\begin{tabular}{lcccccccc}
\hline
\hline
	isotope	&	$^{60}$Co	&	$^{58}$Co	&	$^{57}$Co	&	$^{46}$Sc	&	$^{48}$V	&	$^{54}$Mn	&	$^{59}$Fe	&	$^{56}$Co	\\
\hline																	
half-life / d	&	1925.2	&	70.8	&	271.8	&	83.8	&	16.0	&	312.1	&	44.5	&	77.2	\\
$P$	&	45.5	&	75.6	&	54.8	&	2.3	&	3.7	&	16.5	&	7.2	&	12.4	\\
$\Delta P$	&	26.4	&	35.5	&	19.2	&	0.7	&	0.7	&	7.3	&	5.4	&	5.1	\\
\hline																	
C1: T1	&	1.66	&	3.86	&	6.42	&	0.15	&	0.01	&	1.89	&	0.16	&	0.72	\\
\hspace{0.6 cm}T2	&	1.43	&	0.07	&	2.26	&	0.00	&	0.00	&	0.76	&	0.00	&	0.02	\\
\hdashline
C2: T1	&	1.06	&	4.58	&	4.77	&	0.16	&	0.02	&	1.35	&	0.28	&	0.80	\\
\hspace{0.6 cm}T2	&	0.92	&	0.08	&	1.68	&	0.01	&	0.00	&	0.54	&	0.00	&	0.02	\\
\hdashline
C3: T1	&	1.48	&	5.60	&	6.69	&	0.21	&	0.01	&	1.93	&	0.24	&	1.02	\\
\hspace{0.6 cm}T2	&	1.28	&	0.10	&	2.35	&	0.01	&	0.00	&	0.78	&	0.00	&	0.03	\\
\hdashline
C4: T1	&	1.71	&	6.49	&	7.84	&	0.24	&	0.01	&	2.25	&	0.26	&	1.19	\\
\hspace{0.6 cm}T2	&	1.47	&	0.12	&	2.76	&	0.01	&	0.00	&	0.91	&	0.00	&	0.03	\\	
\hline
\hline
\end{tabular}
\end{footnotesize}
\end{center}
\end{table*}
average production rates from literature are reported in Table~\ref{tab:chap3_cosmicactivation_ge}. The exposure times of the HPGe diodes belonging to C1 to C4 were reconstructed in more detail and used to calculate the activities at the beginning of \textsc{Run-1} and \textsc{Run-2}.
The $^{68}$Ge and $^{65}$Zn associated X-ray line intensities are expected to decrease especially within the initial months of the experiment, but according to background simulations their contribution to the sub-keV$_{ee}$ region is very small (cf.~Section~\ref{Background stability}). Even though not seen by visual inspection of the spectrum, $^{3}$H is predicted to be the third most relevant cosmogenic radioisotope. However, due to its $\beta$-spectrum with endpoint energy of 18.6\,keV and its long half-life of 12.3\,a, it adds a small contribution to the total background and it is practically constant during \textsc{Conus} runs.\\
\indent In Cu, the cosmogenic production of $^{57}$Co, $^{58}$Co, $^{54}$Mn and $^{60}$Co is mainly relevant. To quantify their impact on the sub-keV$_{ee}$ region, the decays  were thoroughly simulated in the Cu parts close to the HPGe diodes (cf.~Section~\ref{Background stability}). It was found that they minimally contribute to the overall background, such that a time-dependent component can be omitted for CE$\nu$NS analysis.

\section{Detector electronics and data acquisition system}
\label{chapter4}

\subsection{Description of electronics circuit}
\label{electronicsCircuit}

\begin{figure*}[ht]
\begin{center}
  \includegraphics[width=0.95\textwidth]{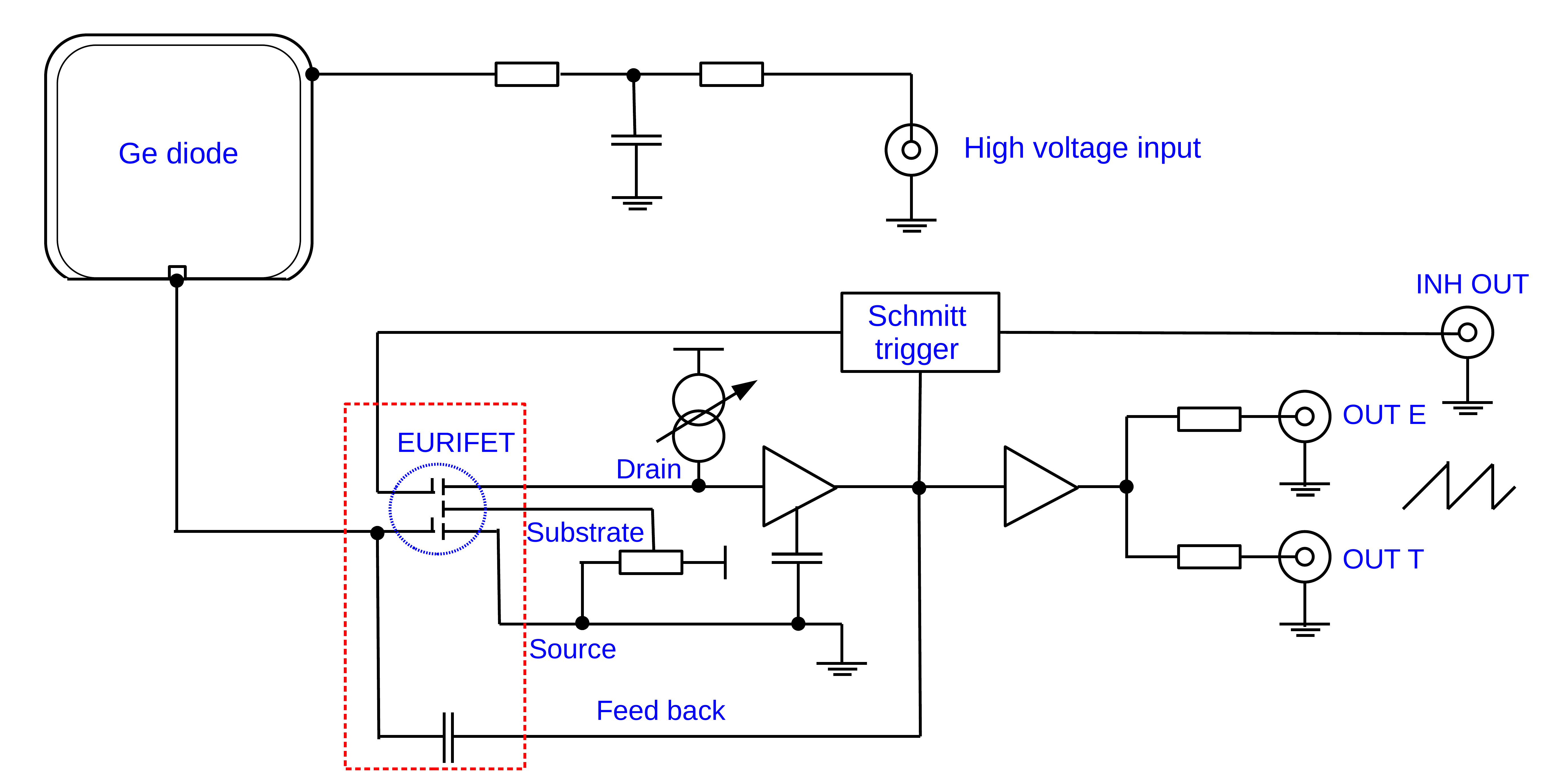}   
  \caption{Simplified representation of the signal detection and amplification electronics chain of the \textsc{Conus} detectors prior digitalization of the raw pulse traces. \textcolor{black}{The cold front-end part of the electronics is outlined by the red dashed box.}}
   \label{fig:ppc_electronics-circuit}  
  \end{center}
\end{figure*}

\begin{figure}[h]
\begin{center}
  \includegraphics[width=0.50\textwidth]{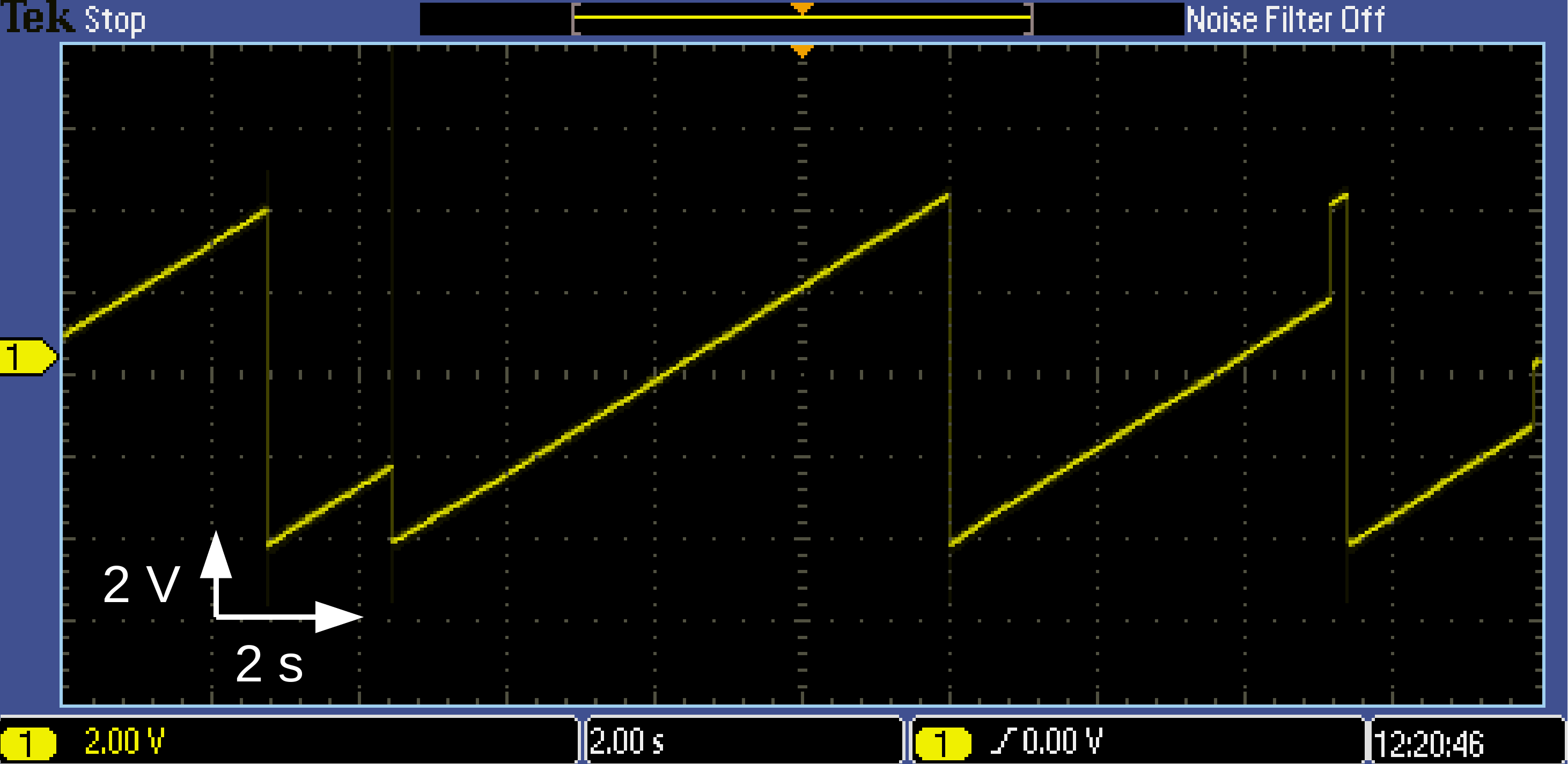}    
  \caption{Exemplary baseline from the pulsed-reset preamplifier of the C4 detector inside the \textsc{Conus} shield at KBR. The dynamic range goes up to $\approx$8\,V, while the time interval between two resets can be large, i.e.\,~up to $\approx$8\,s.}
   \label{fig:pulsed-reset-preamplifier_baseline_c4}  
  \end{center}
\end{figure}

Figure~\ref{fig:ppc_electronics-circuit} depicts the overall signal detection and amplification electronics chain implemented in the \textsc{Conus} detectors.\\
\indent The p-type HPGe diodes are electrically depleted by applying a positive reverse bias voltage of several 1000\,V on the surrounding n+ contact (cf.~Section~\ref{depletion} and Figure~\ref{fig:diodedesign_1}). Ionization events create electron-hole pairs which drift in opposite directions due to the applied potential and the electric field generated by the space  
charge of the depleted HPGe diode. Holes drift to the p+ electrode leading to signal formation.\\
\indent After the p+ contact, the charge signals enter a charge sensitive preamplifier (CSP). Its first amplification stage requires special attention, since it mainly contributes to the overall noise figure. So, starting from the preamplifier model PSC 954-P developed by Mirion-Lingolsheim, a large effort was put into the corresponding cold preamplifier design to find the best compromise between the noise budget, the input capacitance and a high gain. At the same time, new mounting and contacting methods were developed to minimize the stray capacitance seen at the input of the preamplifier. The entire cold preamplifier design uses selected low loss dielectrics and ultra-low radioactive background materials in order to meet simultaneously the ultra-low noise and low background design specifications. To fulfill the latter point, we opted for a custom-built JFET instead of e.g.\,~an application-specific integrated circuit (ASIC). As shown more recently \cite{Barton:2016}, ASICs are indeed very attractive for achieving ultra-low noise levels. They are, however, less reliable in the manufacturing process and more radioactive due to components such as bypass ca\-pa\-ci\-tors needed for ASIC voltage supply.\\
\indent The used DC-coupled CSP includes a pulsed-reset instead of a resistive feedback on the signal contact. This helps to further reduce the noise contribution rising from a potential feedback resistor, however it needs to reset the increasing baselines after saturation of the dynamic range. This is managed by a Schmitt trigger logic. However, under strong detector irradiation and/or in presence of large leakage currents/noise levels, frequent resets can induce a problematic amount of dead time. To overcome this, Mirion-Lingolsheim first managed to achieve very low leakage currents of $\approx$0.25\,pA in all \textsc{Conus} detectors. These can be deduced from the slope of the increasing baselines via $I=dQ/dt=C_{fb}\cdot dV/dt$, with $dV$$\approx$8\,V, $dt$$\approx$8\,s, and $C_{fb}$=0.25\,pF being the internal feedback capacitance (cf.~C4 example in Figure~\ref{fig:pulsed-reset-preamplifier_baseline_c4}). Second, the installation of a massive shield around the HPGe detectors at the \textsc{Conus} experimental site was beneficial not only to create the required low background environment, but also to reduce the pulsed-reset frequency. With reset time windows of [80,~160]\,$\upmu$s selected for the operation of C1-C4 at KBR, the average reset periods turned out to be large, i.e.~around [0.5,~1]\,s, with maxima up to [7,~8]\,s (cf.~Figure~\ref{fig:pulsed-reset-preamplifier_baseline_c4}). So, the dead times induced by resets in the \textsc{Conus} detectors are in general small ($\sim$0.01\%) compared to the contributions from the trigger rate processing time needed by the data acquisition system (DAQ) ([0.1,~4.2]\%; cf.~Section~\ref{Trigger threshold and trigger efficiency}) and from the veto gates ge\-ne\-ra\-ted by the muon ($\mu$) anticoincidence system ([3.5,~5.8]\%).\\
\indent Finally, the CSP generates two equal output signals (energy, time) and a rectangular inhibit signal (TRP), when the Schmitt trigger logic has been activated. The latter is used by the DAQ (cf.~Section~\ref{Lynx DAQ}) to veto all physics and random trigger events generated during the resets of the preamplifier.

\subsection{Data acquisition system}
\label{Lynx DAQ}

For raw data processing we opted for the 32K channel integrated multichannel ana\-lyzer \textsc{Lynx} \cite{Lynx}. For \textsc{Conus}, this DAQ system has several advantages which are summarized here.\\
\indent In terms of space constraints at a reactor site, the \textsc{Lynx} is highly compact offering in one single device high and low voltage power supply and a large set of connectors for different types of inputs: energy, TRP inhibit, HV inhibit and veto signals from the $\mu$-an\-ti\-co\-in\-ci\-dence system \textcolor{black}{(`$\mu$-veto')}. The 10/100 Base-TX Ethernet connection can be used for fast communication with the \textsc{Conus} main PC via TCP/IP.\\
\indent In terms of \textcolor{black}{digitizing and processing signals from raw data \cite{Smith:2003,Knoll:2000},} the \textsc{Lynx} is well suitable for low energy applications, in which small energies and high trigger rates are encountered. It can deal with both types of CSPs, specifically with the pulsed-reset type (cf.~Section~\ref{electronicsCircuit}) used in \textsc{Conus}. Internal gains/attenuators can be adopted/activated, in order to allow e.g.~ an energy range up to 500\,keV$_{ee}$ or a fine-grained binning of $\le$\,1\,eV$_{ee}$ in the CE$\nu$NS ROI. \textcolor{black}{Among a large variety of standard functionalities such as automatic pole/zero cancellation and baseline restorers, it contains a trapezoidal shaping filter (TSF) for energy reconstruction \cite{Jordanov:1994a,Jordanov:1994b}. This offers the possibility to widely adjust the rise time and flat top time of the TSF up to 51\,$\upmu$s and 3.2\,$\upmu$s, respectively. Next to this slow shaping time filter, there is a fast shaping time filter with a short fixed shaping time ($\sim$200\,ns), which allows to identify efficiently pile-up events.}\\
\indent For DAQ control, HV setting and filter manipulation in \textsc{Conus}, we set up a robust interface using the \textsc{Python}-based \cite{Python} software development kit \textsc{Lynx SDK} \cite{Lynx}. This finally allows to save time-stamped energy list modes, which are further analyzed offline with \textsc{Python}- and \textsc{Root}-based \cite{ROOT} scripts.\\
\indent The \textsc{Lynx} DAQ is equipped with an oscilloscope functionality, but does not allow for digitalization of raw CSP traces. This would help to reduce certain types of backgrounds at low energies such as `slow pulses' (cf.~Section~\ref{High purity Ge Diodes}) -- as already demonstrated by other experiments \textcolor{black}{\cite{Aalseth:2011,Aguayo:2013,Li:2014}}. Thus, for the near future \textsc{Conus} pursues the installation of a second DAQ system, which will have the capability to store also the waveform information.

\section{Detector electronic response}
\label{chapter5}

\subsection{Electrical depletion of the HPGe diodes}
\label{depletion}

In order to achieve low leakage currents and thus low noise HPGe diodes, Mirion-Lingolsheim selected p-type HPGe crystals of high quality with proper net impurity concentrations $N_{a-d}:=$ $|N_a$-$N_d|$, with $N_a$ and $N_d$ being the acceptor and donor concentrations. The latter ones guaranteed also the application of lower reverse bias voltages, which are able to electrically fully deplete the HPGe diodes. A multiple approach was applied to deduce these so-called `depletion voltages' (DV) for C1-C4.\\
\indent The manufacturer tracked the C-V and I-V curves in reverse polarity while rising the voltage. A constant capacitance finally confirms the reach of full depletion.\\
\indent The \textsc{Conus} Collaboration remeasured the DVs via an irradiation of the detectors with a $^{57}$Co $\gamma$-ray source and increased the positive voltage in [10,~100]\,V steps. During such a HV scan, the changes in noise on the signal baselines as well as in peak position (PP), peak integral (PI) and energy resolution ($\Delta$E) of the 122.1\,keV $\gamma$-ray line were monitored. A sudden reduction in noise on the signal baseline signalized the attainment of the DV. At full electrical depletion, the PP, PI and $\Delta$E parameters reached constant values within statistics.
\begin{figure}[ht]
	\includegraphics[width=8.5cm]{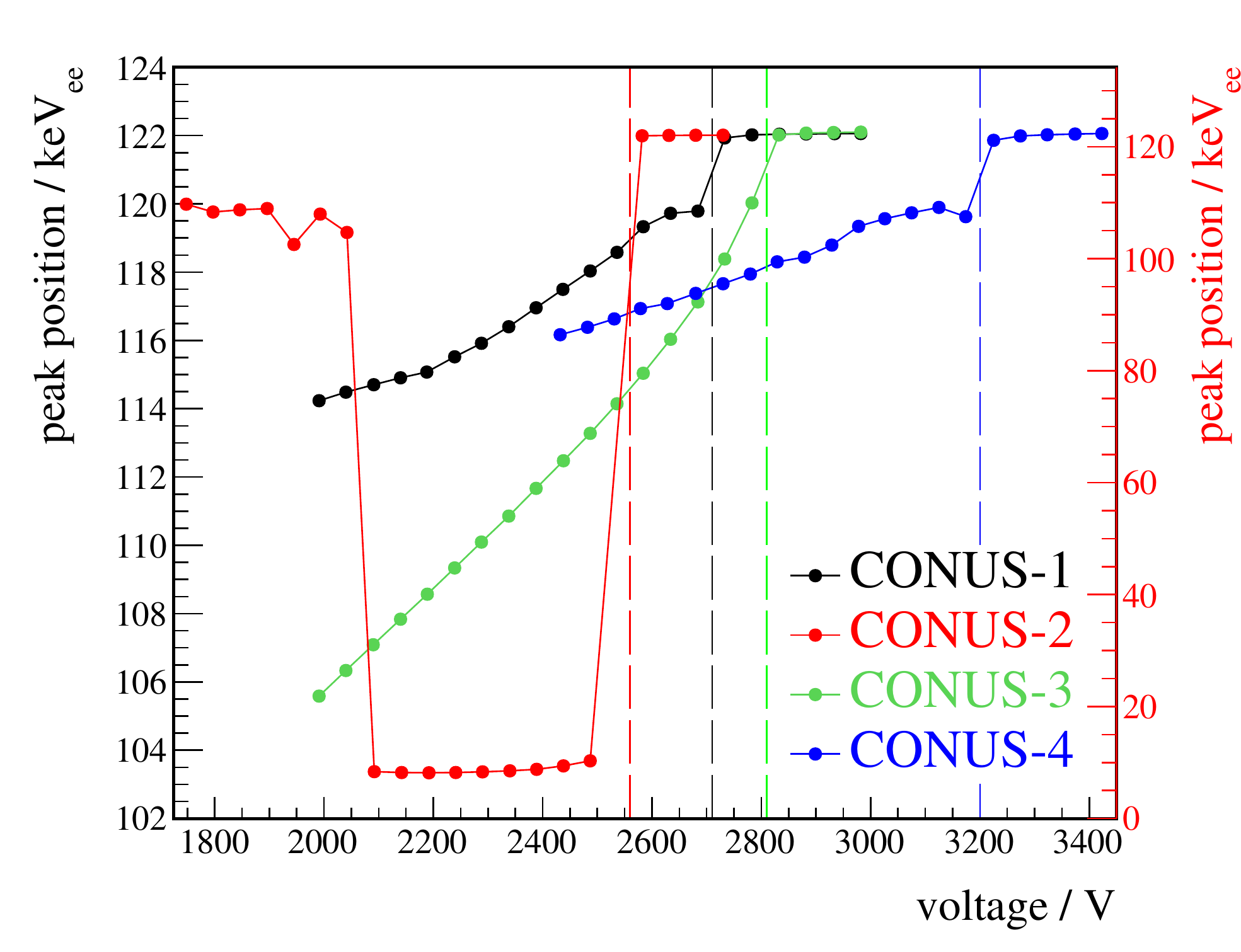}	
	\caption{Peak position of 122.1\,keV $\gamma$-ray line of $^{57}$Co in dependence of the voltage applied at the HPGe diodes of C1-C4. The red y-scale refers to C2 only. The vertical lines mark the depletion voltage determined via the signal baseline method.}
	\label{chapter4:fig_hvconus1}
\end{figure}
\begin{table}[ht]
\begin{center}
\caption{Depletion voltages of the C1-C4 detectors as determined by the \textsc{Conus} Collaboration via two independent procedures: a) visual inspection of the signal baseline on the oscilloscope, b) peak position shift of the 122.1\,keV $\gamma$-ray line. 
\label{chapter4:tab_depletionvoltage}}
\begin{footnotesize}
\begin{tabular}{cccc}
\hline
\hline
detector 	&DV				&DV 				 	&OV \\ 
     		&(baseline)[V] 	&(peak position) [V]	&[V]\\
\hline
C1  		&2730$\pm$1 	&2710$\pm$25           &3000 \\      
C2  		&2070$\pm$1		&2560$\pm$25  			&2700 \\
C3  		&2785$\pm$1		&2810$\pm$25   			&3000 \\
C4  		&3190$\pm$5		&3200$\pm$25     		&3400 \\
\hline
\hline
\end{tabular}
\end{footnotesize}
\end{center}
\end{table}

\begin{figure*}[ht]
\begin{center}
  \includegraphics[width=1.0\textwidth]{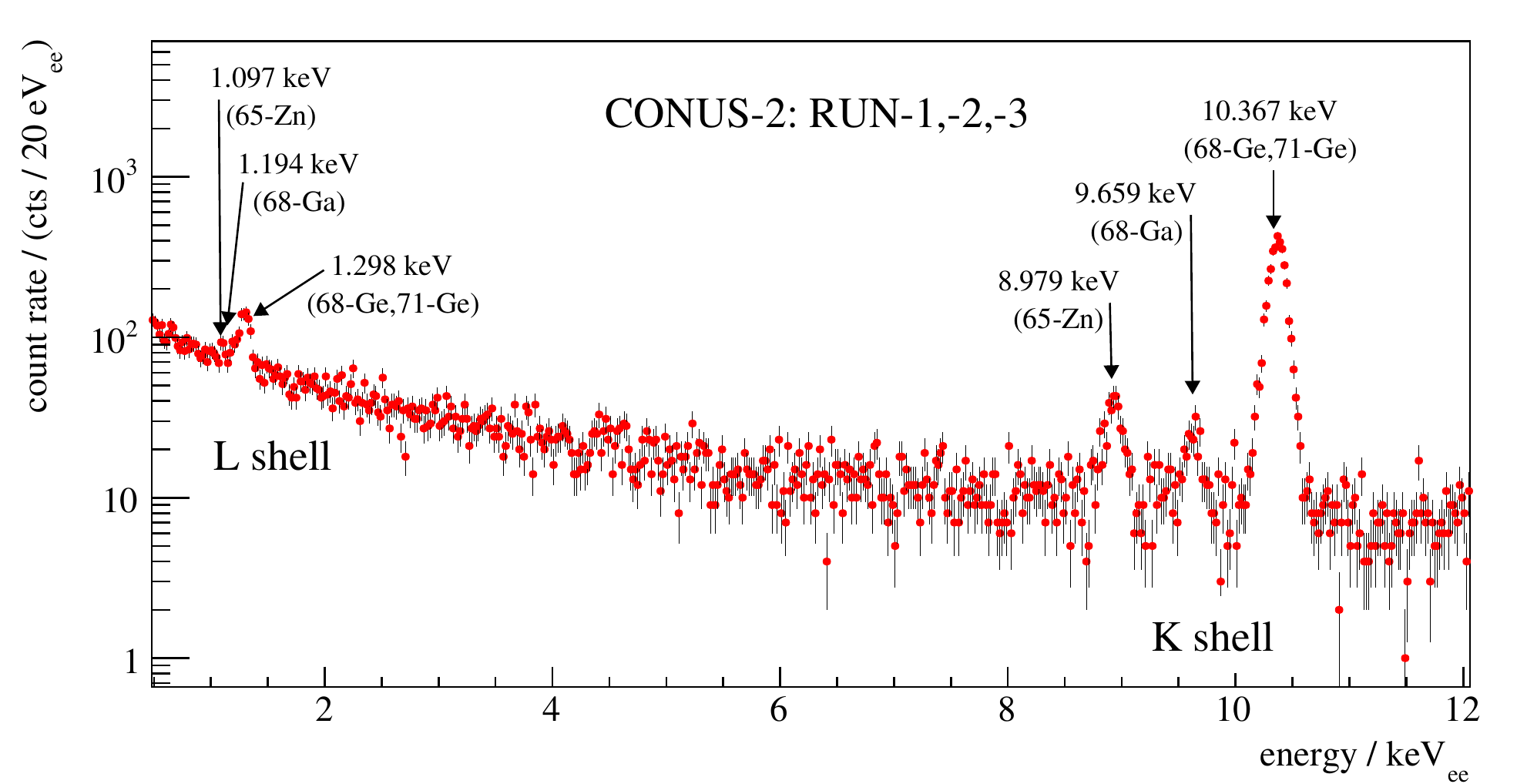}  
   \caption{The ionization energy spectrum of C2, as measured at KBR during the data collection periods \textsc{Run-1} to \textsc{Run-3}. Due to the very low intrinsic and external background only three K shell and corresponding L shell X-ray lines become visible. These are used for energy scale calibration and energy resolution determination.}
   \label{fig:c2_energy-spectrum}  
  \end{center}
\end{figure*}
Moreover, the operational voltages (OV) rec\-om\-men\-ded by Mirion-Lingolsheim are reported.
Figure~\ref{chapter4:fig_hvconus1} depicts the PP curves of C1-C4. The constant plateau in the full depletion regime of the single detectors becomes clearly visible. Next to it, C2 exhibits a broader and deeper discontinuity shortly before reaching DV. This so-called `bubble depletion' \cite{Agostini:2011} or `pinch-off' effect \cite{Abgrall:2014} depends on the detector design and the net impurity concentrations: in the specific case of a PPC HPGe detector, the relevant $N_{a-d}$ values are quoted around the spot-like p+ contact and for the opposite side. For C2 the difference of the two $N_{a-d}$ values turned out to be very small.\\
\indent For the operation of C1-C4 at KBR, we always apply voltages which lie well above the determined DVs. The ones recommended by the manufacturer are at least $\sim$200\,V above the DVs and are also reported in Table~\ref{chapter4:tab_depletionvoltage}.

\subsection{Energy scale: calibration and linearity}
\label{Energy scale: calibration and linearity}

A key requirement for the reduction of the total systematic uncertainty in sub-keV$_{ee}$ physics research is given by a precise energy scale calibration of narrow ROIs. Above energies of a few tens of keV$_{ee}$, we apply the standard method based on an external irradiation of the HPGe detectors with e.g.\,~$^{241}$Am, $^{133}$Ba, $^{57}$Co and $^{228}$Th $\gamma$-ray sources.\\
\indent At lower energies, however, electromagnetic radiation cannot efficiently penetrate the Cu cryostat and the PCCL. Thus, we pursue intrinsic calibrations using X-rays emitted in radioactive decays inside the HPGe diodes. These have further the advantage that the events are generated homogeneously throughout the AV and the reconstructed ionization energy spectrum subse-{\linebreak}quently does not suffer from diode border effects. Specifically, \textsc{Conus} makes use of the X-ray lines corresponding to the binding energies of the K shells at 9.0, 9.7 and 10.4\,keV and of the L shells at 1.1, 1.2 and 1.3\,keV from $^{65}$Zn, $^{68}$Ga and $^{68}$Ge+$^{71}$Ge decays, respectively. A typical energy spectrum -- as measured with C2 at the \textsc{Conus} experimental site during \textsc{Run-1, -2} and \textsc{-3} -- is depicted in Figure~\ref{fig:c2_energy-spectrum}.
\begin{figure}[ht]
\begin{center}
    \includegraphics[width=0.5\textwidth]{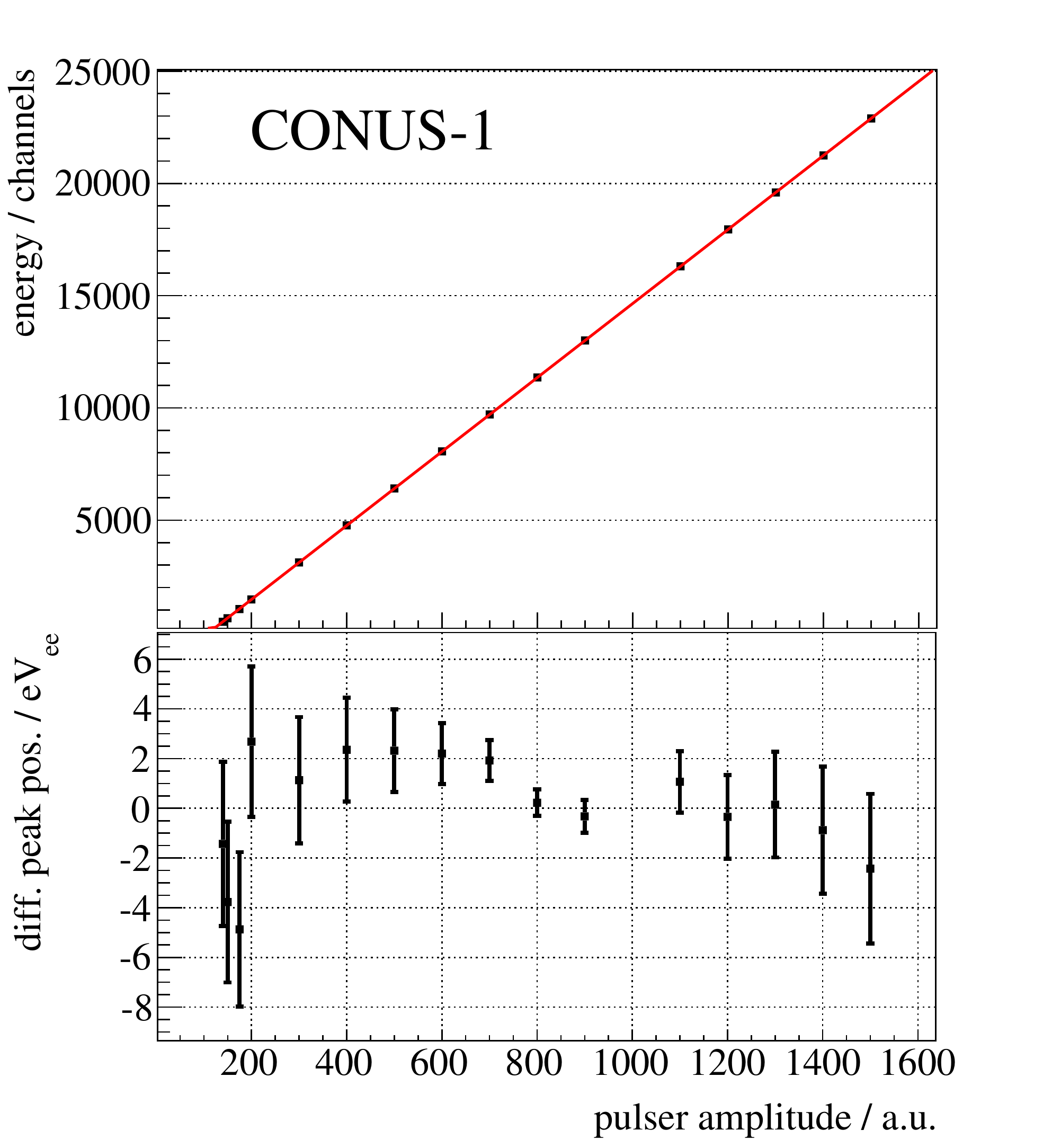}    
   \caption{C1 pulser scan performed during \textsc{Run-1} using the \textsc{Lynx} DAQ. Top canvas: Linearity of the low energy interval [400,~23000]\,channel, corresponding to [0.3,~11]\,keV$_{ee}$. Bottom canvas: Distance of peak positions from linear fit in units of eV$_{ee}$. The shown error bars include the linear fit uncertainties.
   }
   \label{fig:c2_energy-scale-linearity}  
  \end{center}
\end{figure}
With the measured K shell intensities, the weaker L shell intensities can be constrained via well known K/L ratios. For completeness, the binding energies and K/L ratios used in \textsc{Conus} spectral analyzes are reported in Table~\ref{table:x-rays_literature-values}.\\
\begin{table*}[ht]
\begin{center}
\caption{\rm{Literature values of cosmogenic induced X-ray energies and half-lives of Ge-related radioisotopes that are used for the calibration of the \textsc{Conus} detectors. K and L shell values are taken from Ref.~\cite{Bearden-Burr:1967} and the estimated uncertainties from Ref.~\cite{Fuggle-Martensson:1980}, while the K/L ratios are from Ref.~\cite{nucleide.org}.}}
	\vspace{0.3 cm}
	\begin{tabular}{lcccc}
	\hline
	\hline
	nuclide				&half-life 	&K shell [keV]	&L shell [keV]		&K/L ratio\\
	\hline
	$^{68}$Ge			&270.95(26)\,d	&10.3671$\pm$0.0005		&1.2977$\pm$0.0011		&0.1331$\pm$0.0030	\\
	$^{71}$Ge			&11.43(3)\,d	&10.3671$\pm$0.0005		&1.2977$\pm$0.0011		&0.1191$\pm$0.0030	\\
	$^{68}$Ga			&67.83(20)\,min	&9.6586$\pm$0.0006		&1.1936$\pm$0.0009		&0.1109$\pm$0.0013	\\
	$^{65}$Zn			&244.01(5)\,d	&8.9789$\pm$0.0004		&1.0966$\pm$0.0004		&0.1136$\pm$0.0014	\\		
	\hline
	\hline
    \end{tabular}
	\label{table:x-rays_literature-values}	
\end{center}
\end{table*}
\indent The Ge-related radioisotopes are produced con\-ti\-nu\-ously overground via the hadronic component of cosmic radiation (cf.~Section~\ref{Cosmogenic activation in Ge and Cu}). However, during underground storage at a depth below 10\,m w.e., cosmogenic activation via hadrons is already suppressed by three orders of magnitude compared to sea level \cite{Heusser:1995wd,Gilmore:2008}. On the other hand, the still intense cosmic $\mu$-flux at shallow depths can produce $\sim$1\,MeV neutrons \cite{Hakenmueller:2019,Hakenmueller:2016} inside compact Pb shields such as the one used for \textsc{Conus}. These neutrons are able to create $^{71}$Ge atoms inside the HPGe diodes. In the specific case of the \textsc{Conus} site at KBR, the overburden corresponds to 24\,m w.e. and the steady-state production of new $^{71}$Ge atoms leads to a constant 10.4\,keV X-ray line intensity of $\sim$15\,counts kg$^{-1}$d$^{-1}$ in all detectors. With these lines and such count rates, it is possible to calibrate \textsc{Conus} energy spectra based on typical exposures of $>$30\,kg$\cdot$d with a precision of [10,~20]\,eV$_{ee}$ in the ROI. To further improve this, \textsc{Conus} has started a periodic usage of $^{252}$Cf neutron sources (half-life: 2.65\,a). \textcolor{black}{The emitted fast neutrons have energies below 10\,MeV, which are enough to produce short-lived $^{71}$Ge radionuclides, but not long-lived ones such as $^{68}$Ge, $^{65}$Zn or $^{60}$Co (cf. Section~\ref{Cosmogenic activation in Ge and Cu})}. Pretests were done at MPIK with different HPGe detectors \cite{Heusser:2015,Salathe:2015} in 2015. In January 2020, a 0.3\,MBq $^{252}$Cf source with a neutron fluence of (3.3$\pm$0.1)$\times$10$^4$ s$^{-1}$ was deployed next to the \textsc{Conus} shield at KBR for two weeks, leading to a three- to six-fold increase of the 10.4\,keV X-ray line intensities in the C1-C4 detectors. No deterioration of the ener\-gy resolution due to the neutron irradiation was observed. Future $^{252}$Cf source irradiations in \textsc{Conus} are planned, which will allow for a $\leq$5\,eV$_{ee}$ precision on the energy scale calibration in the ROI -- limited mainly by the uncertainties of the literature values.\\
\indent Finally, the DAQ- and detector-dependent energy scale calibration relies on the assumption that it has a highly linear behavior. The validity of this working hypothesis for all detectors was confirmed in dedicated pulse generator (`pulser') scans and with radioactive sources. This is demonstrated exemplarily in Figure~\ref{fig:c2_energy-scale-linearity} for detector C1.

\subsection{Energy resolution}
\label{Energy resolution}

To be sensitive to nuclear recoils producing only a few 100\,eV$_{ee}$ in a 1\,kg massive HPGe detector, extremely low noise levels are required. Based on Mirion-{\linebreak}Lingolsheim's experience, many counter actions in {\linebreak} diode/cryostat/cryocooler/electronics (cf.~Chapter~\ref{chapter2} \linebreak and \ref{chapter4}) construction and assembly were applied in order to meet the \textsc{Conus} experimental specifications.\\
\indent One way to directly characterize the obtained noise level is the energy resolution $\Delta E$ of prominent spectral lines, in the following expressed in terms of full width at half maximum (FWHM). Measurements of injected pulser signals reflect the electronics and environmental noise $\Delta E_{ee}$, while x-/$\gamma$-ray lines contain in addition the statistical fluctuation of the charge release $\Delta E_{sf}$ inside the HPGe diode, and the charge carrier collection efficiency $\Delta E_{cc}$ at the read-out electrode. The energy resolutions were measured within the \textit{Factory Acceptance Test} (FAT) at Lingolsheim, the \textit{Site Acceptance Test} (SAT) at the underground laboratory of MPIK, and at the experimental site of KBR. In all three cases, the environmental noise conditions and energy reconstruction algorithms were first optimized. Herein, we applied a TSF with \textcolor{black}{rise times of [15, 16.8]\,$\upmu$s and flat top times of [0.8, 1.0]\,$\upmu$s. In terms of analogue shaping amplifiers, these values translate into Gaussian shaping times of $\sim$12\,$\upmu$s, which are typical for PPC HPGe detectors of this size.} The energy resolution results are reported in Table~\ref{chapter4:tab_resolution}. The values are well below the design specification limit of $\Delta E_P<85$\,eV$_{ee}$, even under non-laboratory conditions at the KBR site (cf.~Section~\ref{CONUS measurement conditions}). 

\begin{table}[bth]
\begin{center}
\caption{Peak resolutions of the four \textsc{Conus} detectors in terms of FWHM. The first three columns are pulser measurements at Mirion-Lingolsheim, MPIK and KBR. The last two columns correspond to $^{241}$Am and $^{57}$Co source calibrations at MPIK.
\label{chapter4:tab_resolution}}
\begin{footnotesize}
\begin{tabular}{c|ccc|cc}
\hline
\hline
              &\multicolumn{3}{c|}{$\Delta E_P$[eV$_{ee}$] at} & \multicolumn{2}{c}{$\Delta$E[eV$_{ee}$] at} \\
 det.         & Mirion & MPIK  & KBR & 59.6\,keV  & 122.1\,keV     \\
\hline
  C1      	  &	74 & 74$\pm$1 & 69$\pm$1& 327.5$\pm$0.3& 463.2$\pm$0.4  \\
  C2  	      &	75 & 75$\pm$1 & 77$\pm$1& 336.0$\pm$0.4& 491.8$\pm$0.3  \\
  C3  	      &	56 & 59$\pm$1 & 64$\pm$1& 335.5$\pm$0.3& 476.6$\pm$0.5  \\
  C4          &	74 & 74$\pm$1 & 68$\pm$1& 332.6$\pm$0.7& 481.2$\pm$0.4  \\
\hline
\hline 
\end{tabular}
\end{footnotesize}
\end{center}
\end{table}

\subsection{Trigger threshold and trigger efficiency}
\label{Trigger threshold and trigger efficiency}

\begin{figure}[h]
\begin{center}
\vspace{0.3 cm}
\hspace{-0.5 cm}
  \includegraphics[width=.5\textwidth]{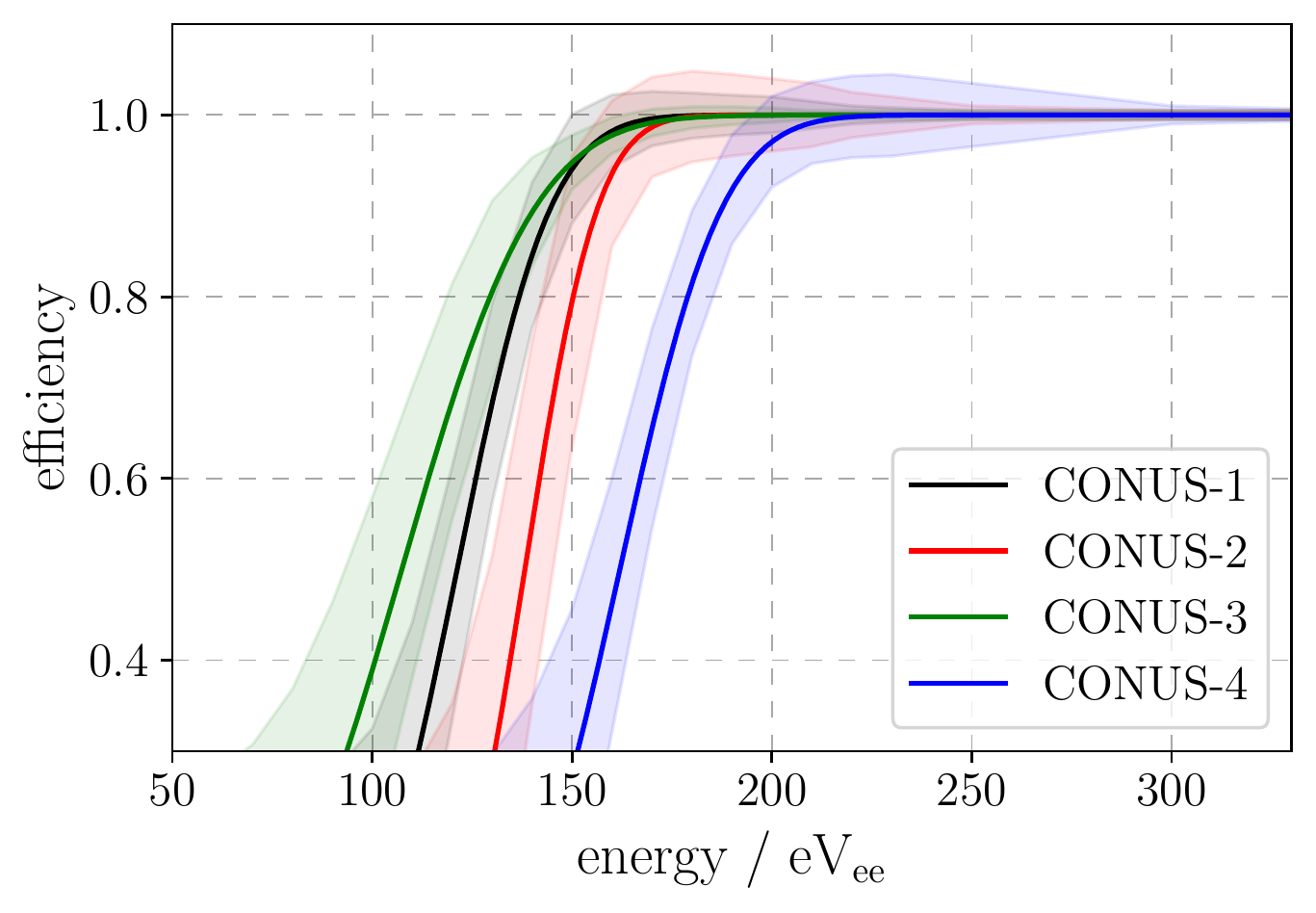}   
  \caption{Trigger efficiency curves of C1-C4, as measured during \textsc{Run-2} of the experiment. Below an efficiency of 50\,\%, the curves are less reliable and thus the shown uncertainty bands might be underestimated.}
  \label{fig:run2_trigger-efficiency-curves}  
  \end{center}
\end{figure}

Starting from the optimized energy filter settings, we tuned the trigger threshold mainly via the \textsc{Lynx} implemented slow discriminator (SD). This defines whether a small step on the baseline -- independent if it is caused by noise or a particle interaction -- is still recognized and digitized as event or is discarded. We selected SD values, \textcolor{black}{which trigger on events with energies above 80-100\,eV$_{ee}$ and} which are a good compromise between the accepted trigger rate (driven by noise and physics event rates) and the induced dead times. The trigger rates are around 100-1000\,Hz and the dead times around {\linebreak} [0.1,~4.2]\%. Table~\ref{chapter4:tab_trgrates_deadtimes} summarizes \textcolor{black}{the chosen SD values as well as} the detector-individual trigger rates and dead times obtained during \textsc{Run-1}, -2 and -3.\\
\indent Below an energy of \textcolor{black}{$\sim$350\,eV$_{ee}$}, noise-induced events start to compete with physical events. At even lower energies \textcolor{black}{around 200\,eV$_{ee}$} the trigger efficiency $\epsilon$ for physical events starts to drop down. \textcolor{black}{The corresponding energy window is called in the following the `noise edge'. The `noise edge' and the region above, in which noise-induced and physical events still coexist, are crucial for the definition of ROIs in CE$\nu$NS searches (see e.g. Ref. \cite{Bonet:2020}).} This emphasizes the importance to explore the detector response of the C1-C4 detectors down to this energy region. A careful examination was done during \textsc{Run-2} of the experiment by means of injected pulses with decreasing amplitudes generated with a \textsc{Tektronics AFG3252} pulser. The trigger efficiency curves are depicted in Figure~\ref{fig:run2_trigger-efficiency-curves}. The energy scale is derived from the energy calibration described in Section~\ref{Energy scale: calibration and linearity} and the experimental measurements were fitted with a Gaussian cumulative distribution function. Characteristic energies E$_{trg}$, at which $\epsilon$ reaches the 50\%, 90\% and 99\% level, are reported in Table~\ref{chapter4:tab_trgeff} for all four detectors. For \textsc{Run-3} the values are similar. As one can see, $\epsilon>$\,99\% persists down to [165,~210]\,eV$_{ee}$, which coincides with 2.2-3.1 times the pulser resolutions $\Delta E_P$ (cf.~Section~\ref{Energy resolution}). During \textsc{Run-1} a pulser scan with an \textsc{EG{\&}G Ortec 448} pulser was performed. Due to periodicity instabilities of this device it was not possible to determine the efficiency curve well below 200\,eV$_{ee}$, but above this energy $\epsilon>$\,95\% was confirmed.

\begin{table}[bth]
\begin{center}
\caption{Slow discriminator values SD, accepted trigger rates $R_{trg}$ and $R_{trg}$-induced dead times $t_{d}$ in \textsc{Run-1, -2} and \textsc{-3} of the \textsc{Conus} experiment, reported separately for each detector.
\label{chapter4:tab_trgrates_deadtimes}}
\begin{footnotesize}
\begin{tabular}{c|ccc|ccc}
\hline
\hline
            &\multicolumn{3}{c|}{\textsc{Run-1}} & \multicolumn{3}{c}{\textsc{Run-2 \& -3}} \\
 det.       &SD  & $t_{d}$  &  $R_{trg}$ &SD & $t_{d}$  &  $R_{trg}$  \\
            &[\%] &[\%] &[Hz] &[\%] &[\%] &[Hz]  \\
\hline
C1			&2.4	&1.3	&300	&[2,~2.1]	&[3.7,~4.2]   	&1500\\
C2			&3.2	&1.8	&125	&[2,~2.6]	&[1.0,~4.0]		&$\geq$325\\
C3			&2.4	&0.3	&150	&[1.8,~2.6]	&[1.5,~2.0]    	&$\geq$625\\
C4			&2.6	&0.5	&275	&[2.2,~2.8]	&[0.1,~2.1]		&$\geq$125\\  
\hline 
\hline
\end{tabular}
\end{footnotesize}
\end{center}
\end{table} 

\begin{table}[bth]
\begin{center}
\caption{Energy values E$_{trg}$ of \textsc{Conus} detectors, at which the trigger efficiency $\epsilon$ reaches characteristic values. The ratios of E$_{trg}$($\epsilon$=99\%) over $\Delta E_P$ are reported as well.  
\label{chapter4:tab_trgeff}}
\begin{footnotesize}
\begin{tabular}{c|ccc|c}
\hline
\hline
    	&\multicolumn{3}{c|}{E$_{trg}$ [eV$_{ee}$] for:} & E$_{trg}$/$\Delta E_P$ \\
det. 	&$\epsilon$=50\% &$\epsilon$=90\% &$\epsilon$=99\%  &(for $\epsilon$=99\%)		    \\
\hline
C1		&121	&145	&165	&2.4			\\
C2		&138	&156	&172	&2.2			\\
C3		&108	&141	&169	&2.6			\\
C4		&162	&188	&210	&3.1			\\
\hline 
\hline
\end{tabular}
\end{footnotesize}
\end{center}
\end{table}

\subsection{Rejection of noise and spurious events}
\label{Rejection of noise and spurious events}

\begin{figure}[h]
\begin{center}
  \includegraphics[width=.5\textwidth]{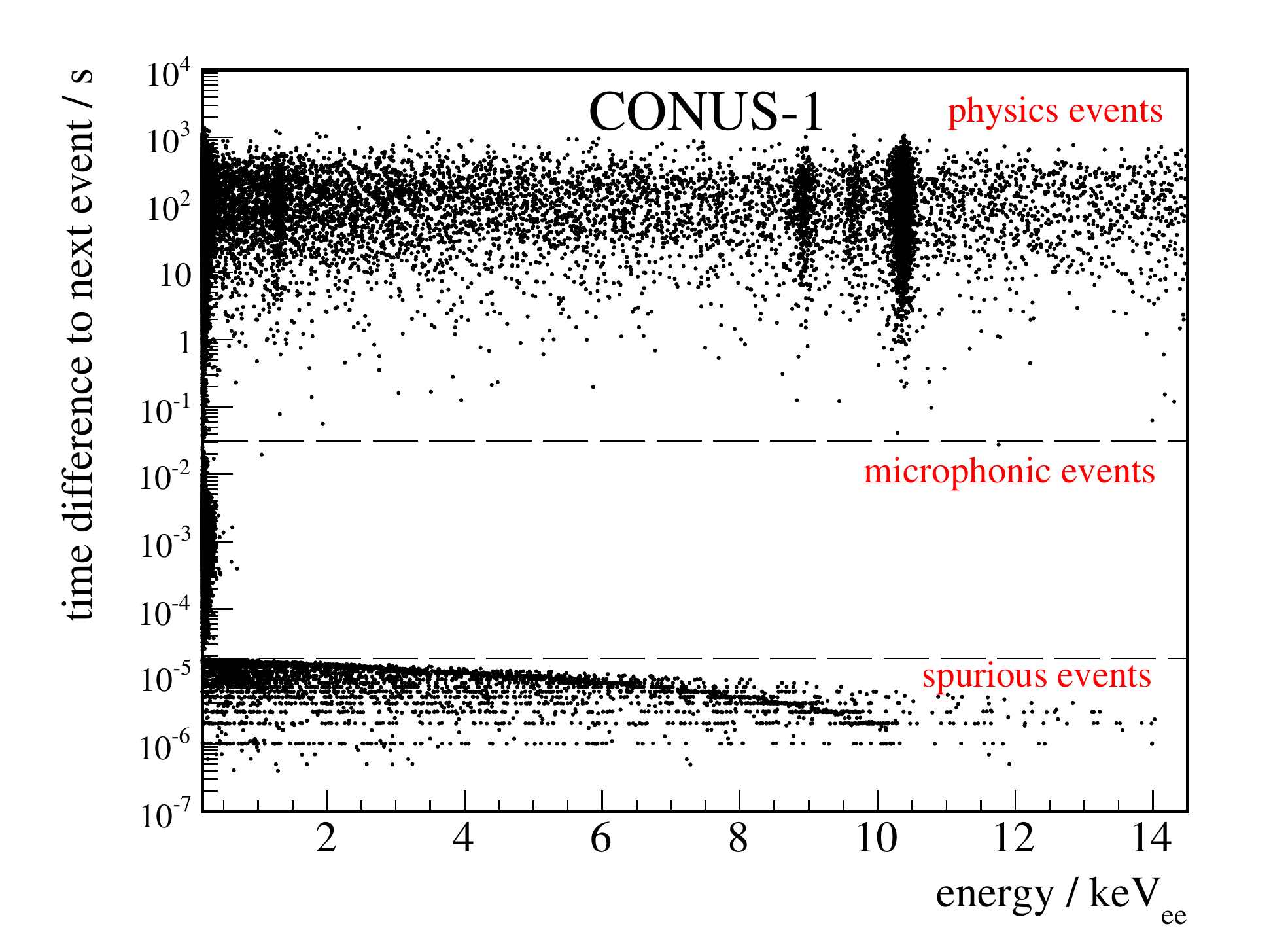}      
   \caption{Time difference distribution as function of ener\-gy, as measured with C1 during \textsc{Run-1}. Two dotted lines at $\Delta t$=0.03\,s (10$^{-1.5}$\,s) and $\Delta t$=25\,$\upmu$s (10$^{-4.6}$\,s) separate three populations of events. The first one is related to physical events. The intermediate one is mainly attributed to microphonic events. The fastest one is mostly connected to spurious pile-up events, which were not correctly reconstructed by the DAQ.}
   \label{fig:timedifference_vs_energy_conus1_reactorON_run1}  
  \end{center}
\end{figure}

The low energy range just above the `noise edge' can be further optimized by offline data analyses. In the case of \textsc{Conus}, the study of the time difference $\Delta t$ bet\-ween single events turned out to be a useful tool to investigate and discriminate noise as well as spurious events. To visualize these populations, the time difference distribution (TDD) was plotted against energy for events above a detector-dependent cut threshold, which \textcolor{black}{exemplary for the C1 detector (\textsc{Run-1})} lies slightly above 170\,eV$_{ee}$. As one can see in Figure~\ref{fig:timedifference_vs_energy_conus1_reactorON_run1} for this detector, three categories were identified:\\
\indent First, events with $\Delta t >$\,0.03\,s are Poisson-distributed, as expected from randomly distributed physics events. Characteristic Ge X-ray lines (cf.~Table~\ref{table:x-rays_literature-values}) show up at the corresponding energies. \textcolor{black}{In general, the detector- and run-dependent percentage of these events to the total number is in the range of [60, $\sim$100]\%.}\\
\indent Second, events with 25\,$\upmu$s $< \Delta t <$\,0.03\,s and a frequency maximum around 1\,kHz are mostly due to microphonic noise. \textcolor{black}{The detector- and run-dependent contribution to all events above the cut threshold is between [1, 30]\%.} Microphonic events are produced by mechanical vibrations originating from the cryocooler or from the reactor environment. They can be continuous or last only a few seconds. Sequences involving up to a few 100 events within less than 1\,s were sometimes encountered during reactor OFF times, in a few occasions also during reactor ON periods, e.g at the time when the main cooling pumps of the reactor are turned on. In the ionization energy spectrum microphonic events are dispersed above the `noise edge' and could pollute the ROI \textcolor{black}{(cf. Ref. \cite{Bonet:2020})}. However, since the first and second category are well separated, it is possible to apply a TDD cut to reject this kind of microphonic events with an efficiency of nearly 100\%.\\
\indent Third, fast sequences of two or more single events were observed with $\Delta t <$\,25\,$\upmu$s. \textcolor{black}{The detector- and run-dependent contribution to all events is around [1, 10]\%. However, the intensity of these fast events} can be influenced by the fast discriminator (cf.~Section~\ref{Lynx DAQ}), indicating that the \textsc{Lynx} built-in pile-up rejector fails in this particular situation. It turned out that the energy of the last events of such sequences is reconstructed correctly, while not that of the other events. Thus, these events are not trustful and commonly rejected in \textsc{Conus} data analyses.

\section{Long-term detector stability}
\label{chapter6}

\subsection{CONUS measurement conditions}
\label{CONUS measurement conditions}

Nuclear power plants are ideal for CE$\nu$NS research, since the reactor cores provide a highly intense, but also variable $\bar{\nu}$-flux, and they can be approximated as a point-like source for typical distances of the experimental setups. During reactor ON periods the $\bar{\nu}$-flux/spec\-trum varies due to changes in the isotopic composition with the increasing burn-up of the nuclear fuel. Additionally, nuclear power plants in Germany are highly engaged in load follow operations. For KBR, the possible range is from 100\% down to approximately 60\% nominal power. Reactor OFF periods at KBR cover regular outages, occurring typically once a year and lasting up to four weeks, and shut-down periods due to maintenance work on short notice. Since the \textsc{Conus} setup installation in January 2018, there were two outages (March 31-May 7, 2018; June 8-July 9, 2019) and two shorter shut-down periods (April 13-24, 2019; November 17-23, 2019).\\
\indent However, in order to compare the energy spectra collected during variable $\bar{\nu}$-flux periods, detectors and ex\-pe\-ri\-men\-tal sites have to guarantee stable measurement conditions, i.e.\,~constant background and noise levels. In close proximity to reactor cores (cf.~Section~3.1 in Ref.~\cite{Hakenmueller:2019}), these requirements are {\it{a priori}} not easily fulfilled. Table~\ref{chapter6:compare-underground-locations} summarizes typical measurement conditions for the \textsc{Conus} detectors encountered during commissioning at the MPIK underground laboratory and during operation at the experimental site, i.e.~room A408 in the KBR reactor building. Especially challenging at the reactor site might be a $P_{th}$ correlated radiation background (e.g.\,~from escaping fission neutrons or $\gamma$-radiation from $^{16}$N decays in the primary cooling cycles \cite{Hakenmueller:2019}) or noise (e.g.\,~from the steam generators), which could mimic CE$\nu$NS signals. Moreover, unprecedented situations might occur. In the case of \textsc{Conus}, there were a few of them during outage periods. These include short, but strong increases in room temperature (up to 31\,\textdegree{}C) and enhanced radon (Rn) concentrations in air. On July 5-6, 2019, a regular leakage test of the reactor safety vessel was performed. For this purpose, the air pressure inside the entire containment was set to 1.5\,bar absolute pressure. To avoid damage to the \textsc{Conus} detectors, their cryostats had to be ventilated and filled with argon of 99.9999\% purity before the test, and evacuated afterwards.\\
\indent Within the limitations at the reactor site, we addressed all points in Table~\ref{chapter6:compare-underground-locations}. I.e.\,~an elaborated shield (cf. Section~4.1.2 in Ref.~\cite{Hakenmueller:2019}) was constructed, optimized and commissioned at MPIK prior to its installation at KBR. Relevant parameters at KBR site were carefully monitored, environmental conditions improved, background/noise sources mitigated or quantified via independent measurements. MC simulations helped to understand the impact of the remaining contributions on the \textsc{Conus} energy spectra. The associated long-term studies described in Sections~\ref{Energy scale and energy resolution stability}, \ref{Detector noise stability} and \ref{Background stability} led to a selection of measurement periods, in which the electronic response, the noise and background levels of the \textsc{Conus} detectors are stable enough to guarantee a robust CE$\nu$NS research.

\begin{table}[bth]
\begin{center}
\caption{Measurement conditions and typical environmental parameters at the MPIK underground laboratory \cite{Heusser:2015,Heusser:1993} and at the \textsc{Conus} experimental site at KBR \cite{Hakenmueller:2019}. 
\label{chapter6:compare-underground-locations}}
\begin{footnotesize}
\begin{tabular}{lcc}
\hline
\hline
parameter								&underground   		&room A408                 \\
 					        			&lab. at MPIK       &at KBR                    \\
\hline
overburden [m w.e.]         			&15            		&24                        \\
$\mu$-flux reduction          			&$\sim$[2,~3]       &$\sim$[3,~4.5] $^{a}$        \\
Rn conc. [Bq m$^{-3}$]  				&60            		&175$\pm$35 $^{b}$  \\
Rn mitigation          					&N$_{2}$ flush 		&CAB flush $^{c}$  \\
neutrons [cm$^{-2}$d$^{-1}$GW$^{-1}$]	&natural            &191$\pm$8         \\
$^{16}$N $\gamma$s [cm$^{-2}$d$^{-1}$GW$^{-1}$] &negligible	&$\mathcal{O}$: 10$^6$ $^{d}$      \\
floor contamination         			&natural       		&$^{60}$Co, $^{137}$Cs\\
room temperature [\textdegree{}C]     	&23$\pm$1       	&26$\pm$3; 20$\pm$2 $^{e}$   \\
access                      			&24/7          		&10/5  \\
remote control              			&available     		&none  \\
\hline 
\hline
\end{tabular}
\end{footnotesize}
\end{center}
\scriptsize{$^{a}$ Relative $\mu$-flux reduction at KBR cmp. to MPIK: 1.62$\times$.}\\
\scriptsize{$^{b}$ Average value from \textsc{Run-1} and \textsc{Run-2} is reported.}\\
\scriptsize{$^{c}$ Compressed air bottles (CAB) stored for $>$3 weeks before usage.}\\
\scriptsize{$^{d}$ Local flux depends on the solid angle and on the position relative to the \textsc{Conus} shield.}\\
\scriptsize{$^{e}$ Average \textsc{Run-1} and \textsc{Run-2} values are reported separately. In \textsc{Run-2} the room temperature was lowered via an air conditioner system installed inside the tent surrounding the \textsc{Conus} shield.}\\
\end{table}

\subsection{Peak position and energy resolution stability}
\label{Energy scale and energy resolution stability}

The PP and $\Delta$E stabilities of the \textsc{Conus} detectors were determined in three different ways during the experimental runs at KBR. These comprise pulser and $^{228}$Th source measurements as well as a continuous monitoring of the prominent X-ray line at 10.4\,keV. The first two measurement types were repeated on an almost daily basis during reactor outages and every two to four weeks during reactor ON periods. Each of these measurements lasted typically 10\,min (100\,Hz pulse frequency) and 2\,h (15\,kBq $^{228}$Th source placed at 20\,cm distance from the PPC HPGe diodes), respectively, in order to achieve a fit uncertainty on the PP and $\Delta$E values, which are in the range of [4,~6]\,eV$_{ee}$. Due to the modest intensity of the 10.4\,keV X-ray line, time-bins of one month were selected to allow again for a PP fit uncertainty of $\sim$5\,eV$_{ee}$.\\
\indent Table~\ref{chapter4:tab_longtermstability_pp_eneres} summarizes the standard deviations $\sigma$ of the PP and $\Delta$E mean values of the three type of peaks for the \textsc{Run-1} and \textsc{Run-2} time periods. These are in general very small and did not deteriorate compared to tests previously performed under laboratory conditions at MPIK. Especially, the $\sigma_{PP}$ values of [2,~9]\,eV$_{ee}$ from the 10.4\,keV X-ray line and of [1,~8]\,eV$_{ee}$ from the pulser peaks demonstrate that the energy scale of the four \textsc{Conus} detectors at low energies remained very stable over time periods of several months.

\begin{table*}[btp]
\begin{center}
\begin{footnotesize}
\begin{tabular}{c|cc|cc|cc}
\hline
\hline
detector & \multicolumn{2}{c|}{10.4\,keV} & \multicolumn{2}{c|}{238.6\,keV} & \multicolumn{2}{c}{pulser at (320\,eV; 430\,eV)}\\
& $\sigma_{PP}$  & $\sigma_{FWHM}$  & $\sigma_{PP}$  & $\sigma_{FWHM}$  & $\sigma_{PP}$  & $\sigma_{FWHM}$ \\
\hline
C1	&2.3; 5.4  &	10.5; 14.7	&	16.6; 30.3	&	9.2; 5.0	&	4.0; 0.6	&	1.6; 3.2\\
C2	&4.9; 9.0  &	15.2; 21.1	&	15.9; 13.6	&	16.0; 10.6	&	6.0; 2.0	&	3.0; 3.1\\
C3	&3.7; 2.9  &	15.7; 29.7	&	21.3; 36.5	&	12.6; 8.2	&	6.9; 7.8	&	1.4; 2.6\\
C4	&2.3; 5.5  &	10.8; 12.5	&	13.7; 19.3	&	11.3; 10.8	&	4.7; 0.7	&	1.6; 4.7\\
\hline               
\hline
\end{tabular}
\end{footnotesize}
\caption{Standard deviations (unit: eV$_{ee}$) of PP and $\Delta$E of the 10.4\,keV X-ray and 238.6\,keV $\gamma$-ray line as well as of the pulser peak at (320\,eV; 430\,eV) during the described measurement periods (\textsc{Run-1}; \textsc{Run-2}) of the \textsc{Conus} experiment.
\label{chapter4:tab_longtermstability_pp_eneres}}
\end{center}
\end{table*}

\subsection{Detector noise stability} 
\label{Detector noise stability} 

\begin{figure}[h]
\begin{center}
\includegraphics[width=.5\textwidth]{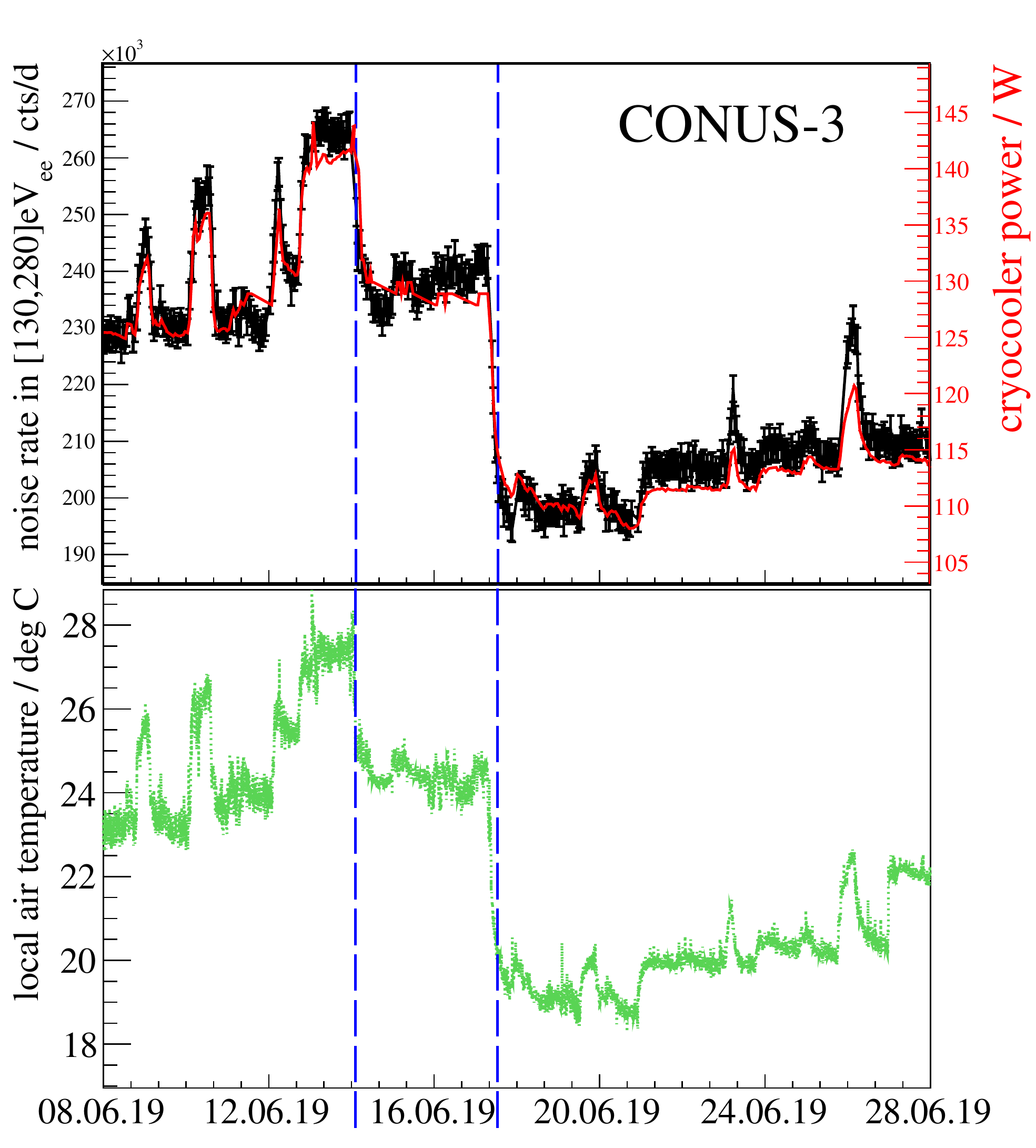}  
\caption{C3 detector during the outage of \textsc{Run-2}: The local air temperature, the cryocooler power consumption and the resulting noise integral in [130, 280]\,eV$_{ee}$ are shown. They are strongly correlated. The two dotted blue lines reflect the installation and optimization of the so-called `cold-house', which surrounds the \textsc{Conus} setup in room A408 of KBR.}
\label{fig:KBR_cryocooler_vs_noiseintegralpart_c3_run2}  
\end{center}
\end{figure}

Since the `noise edge' (cf.~Section~\ref{Trigger threshold and trigger efficiency}) is close to the CE$\nu$NS ROI \textcolor{black}{(cf. Section \ref{chapter1} and Ref. \cite{Bonet:2020})}, we precisely monitored the stability of the noise rate $R_n$ below the ROI and in\-ve\-sti\-gated potential correlations between noise variations and external parameters.\\
\indent First, an intense and well matching correlation of $R_n$ with the cryocooler power consumption $P_{c}$ and thus with its work load was observed. Two effects can cause a change in $P_{c}$. It can occur if the vacuum inside a cryostat slowly deteriorates in time. A small decrease of the PPC HPGe diode temperature $T_d$ by a few degrees (cf. Section~\ref{Cryocooler}) can help, but a cryostat pumping is recommended. Among the \textsc{Conus} detectors, C2 turned out to suffer from such a vacuum instability and was pumped onsite in February 2019. Further, $P_{c}$ is strongly correlated to the local air temperature $T_{l}$ around the fan-in ventilators of the cryocoolers. The reaction of the cryocoolers on $T_{l}$ variations is instantaneous. The $P_{c}$ curve of the C3 cryocooler is depicted in Figure~\ref{fig:KBR_cryocooler_vs_noiseintegralpart_c3_run2} for the outage period in 2019 together with the correlated $T_{l}$ and $R_{n}$ measurements. In the initial phase of the outage from June 8 until 14, 2019, the tent surrounding the \textsc{Conus} setup has not yet been fully closed. So, the $R_n$ values in [130,~280]\,eV$_{ee}$ increased with the rising and fluctuating overall room temperature $T_{r}$. As observed for C3, a 1\,\textdegree{}C increase in $T_{l}$ enhances $P_{c}$ and $R_n$ by 3 and 4\%, respectively. Between June 14-18, 2019, the \textsc{Conus} tent was fully closed and connected to an air conditioner system. Since then, this `cold house' has allowed to lower and stabilize $T_{l}$ around [18,~20]\,\textdegree{}C. Next to this measure, we applied a noise-temperature correlation (NTC) cut offline in all previously collected \textsc{Conus} data, which considers only time periods with small $R_n$ fluctuations below [5,~10]\% within a given time period. This led to a larger loss of exposure, but also to datasets with very stable noise rates above 200\,eV$_{ee}$. This benchmark is used as one criterion to define the lower limit of CE$\nu$NS ROIs in \textsc{Conus} analyses \textcolor{black}{(cf. Ref.~\cite{Bonet:2020})}.\\
\indent Beside the cryocooler induced noise, we inquired the potential existence of a noise component correlated to the reactor $P_{th}$. Natural candidates could be mechanical vibrations induced e.g.~by the reactor pumps of the primary cooling cycle or the steam generators during reactor ON periods. With the help of a seismograph installed in room A408, the noise in a large frequency band was registered. A first conclusion was that the seismic activity reflects the reactor operations, but these are not fully aligned with the reactor $P_{th}$. A full discussion about this topic will follow in a separate publication. Surprisingly, a $P_{th}$ correlated noise component was found in C3 and to some extent in C2 during short periods in \textsc{Run-1} and bet\-ween \textsc{Run-1} and \textsc{Run-2}. 
It is mainly induced by a cross talk with the \textsc{Conus} \textcolor{black}{$\mu$-veto}. The corresponding scintillator plates are integrated in the outer layers of the \textsc{Conus} shield (cf. Figure~9 in \cite{Hakenmueller:2019}), which are less protected against the $P_{th}$ correlated $^{16}$N $\gamma$-radiation onsite. Fortunately, it was shown that this noise populates only the `noise edge' and that its contribution to the CE$\nu$NS ROIs is negligible.

\subsection{Background stability}
\label{Background stability}

The spectral background registered by the \textsc{Conus} detectors was investigated in detail with a full MC-based background decomposition \cite{2020:Hakenmueller} and with the help of auxiliary on-/offsite measurements. 
\begin{table*}[ht]
\begin{center}
\caption{\textcolor{black}{Relative contributions $R_b$ (unit: \%) from different background sources to the measured total background rates}, as determined for the C1-C3 detectors at the beginning of \textsc{Run-1}. Typical relative rate changes are reported in the last column.
\label{chapter4:tab_backgroundrates}}
\begin{footnotesize}
\begin{tabular}{l|ccc|ccc|c}
\hline
\hline
type 	      	&\multicolumn{3}{c|}{$R_b$ [\%] in [0.4,~1.0]\,keV$_{ee}$}    &\multicolumn{3}{c|}{$R_b$ [\%] in [2,~7]\,keV$_{ee}$}& $\Delta R_b/R_b$\\ 
				&C1&C2&C3&C1&C2&C3&[\%]\\ 
\hline
prompt $\mu$-ind. signals \textcolor{black}{(with $\mu$-veto)}	&33.4	&52.5	&33.8 	 &32.3 	&46.9  	&35.3   	&1\,a: $\le$2 \\
delayed $\mu$-ind. Ge isomers  			&0.4 	&0.5 	&0.4 	 &0.3 	&0.3 	&0.3  		&1\,a: $\le$2 \\
$\mu$-ind. neutrons in concrete			&6.3	&10.3 	&6.3 	 &4.4  	&6.5 	&5.1    	&1\,a: $\le$2 \\
cosmogenic radioisotopes 				&3.2 	&3.6	&1.8 	 &4.7  	&6.1   	&2.5    	&1\,a: 30-40 \\
$^{210}$Pb in shield 					&3.7 	&1.5	&0.8 	 &5.5  	&2.2   	&1.3 	 	& --\\
$^{210}$Pb inside cryostat				&52.7 	&27.8 	&50.7 	 &52.4 	&31.8  	&42.6   	& --\\
airborne Rn   							&0.2 	&3.6	&6.1 	 &0.3 	&6.1   	&12.8   	&0-100\\
fission neutrons / $^{16}$N $\gamma$s 	&0.1 	&0.2 	&0.1   	 &$<$0.1&$<$0.1 &$<$0.1 	&0-100\\
\hline
\hline
\textbf{total [counts kg$^{-1}$d$^{-1}$]}	&11.5$\pm$0.9&6.4$\pm$0.8 &13.1$\pm$1.0 &36.3$\pm$1.6 &23.5$\pm$1.3 &32.4$\pm$1.7 & \\
\hline
\hline
\end{tabular}
\end{footnotesize}
\end{center}
\end{table*}
A full background discussion goes beyond the scope of this report and is postponed to a future publication. This section focuses on the average rates of different background contributions and their stability. Table~\ref{chapter4:tab_backgroundrates} summarizes the background rates measured in two energy windows of the C1-C3 detectors at the beginning of \textsc{Run-1}: the region $E_1$=[0.4,~1.0]\,keV$_{ee}$ includes part of the CE$\nu$NS ROIs, and $E_2$=[2,~7]\,keV$_{ee}$ the background between the L and K shell X-ray lines (cf.~Figure~\ref{fig:c2_energy-spectrum}).\\
\indent The first background class is related to prompt and delayed $\mu$-induced signals generated inside the shield. Compared to all other contributions, prompt $\mu$-induced signals represent the most intense background component, i.e.\,~(160$\pm$10) and (430$\pm$15) counts kg$^{-1}$d$^{-1}$ in $E_1$ and $E_2$, respectively. \textcolor{black}{If the $\mu$-veto is applied, these rates are reduced by $\sim$97\% down to 5 and 13 counts kg$^{-1}$d$^{-1}$.} Furthermore, $\mu$-induced neutron interactions in Ge can lead to delayed isomeric states, whose de-excitation time can exceed the applied gate length of 410\,$\upmu$s of the $\mu$-veto. However, their contribution to the background rate is modest, i.e.\,~(0.08$\pm$0.01) counts kg$^{-1}$d$^{-1}$ in $E_1$ and (0.10$\pm$0.01) counts kg$^{-1}$d$^{-1}$ in $E_2$. The existence of a seasonal $\mu$-modulation due to temperature and density changes in the atmosphere has not yet been confirmed for the shallow depth \textsc{Conus} experimental site, but is expected to be $\le$2\% \cite{2012:Cecchini,2017:Abrahao,2017:Arunbabu}.\\
\indent A second class is given by the decays of the cosmogenic radioisotopes (cf.~Section~\ref{Cosmogenic activation in Ge and Cu}). Their contribution to $E_1$ and $E_2$ vary between the detectors around [0.2,~0.4] counts kg$^{-1}$d$^{-1}$ and [0.6,~1.7] counts kg$^{-1}$d$^{-1}$ at the beginning of \textsc{Run-1}. The integral contribution is mainly caused by $^{71}$Ge, $^{68}$Ge and $^{3}$H decays. Depending on the detector activation histories and \textit{in-situ} cosmic production of $^{71}$Ge, the cosmogenic contribution is expected to drop down to 60-70\% within the first year of data collection and to stabilize around 30-40\% within a few years.\\ 
\indent The \textsc{Conus} shield as well as the \textsc{Conus} detectors contain traces of the $^{210}$Pb radioisotope, whose decays lead to a third relevant background contribution. A MC-based calculation allowed to extract a detector-dependent rate of [2,~7] counts kg$^{-1}$d$^{-1}$ in $E_1$ and [8,~21] counts kg$^{-1}$d$^{-1}$ in $E_2$. With a half-life of 22.3\,a, this $^{210}$Pb-induced background can be con\-si\-de\-red as constant during a typical \textsc{Conus} run.\\
\indent A fourth class is related to decays of airborne Rn. Even though the inner-most detector chamber of the \textsc{Conus} shield is continuously flushed with Rn-free air, Rn can sporadically appear (in terms of the 351.9\,keV $\gamma$-ray line) due to e.g.~a non-sufficient flushing. With the help of MC simulations the impact of $^{222}$Rn and its progenies on the low energy spectrum was simulated. The maximum calculated rates are 0.9 counts kg$^{-1}$d$^{-1}$ in $E_1$ and 4.7 counts kg$^{-1}$d$^{-1}$ in $E_2$.\\
\indent A fifth background class consists of fission neutrons and $^{16}$N $\gamma$-rays, since they are fully $P_{th}$ correlated. As already demonstrated in Ref.~\cite{Hakenmueller:2019}, their impact on the very low energy window is very small, i.e.\linebreak	(0.010$\pm$0.005)\,counts kg$^{-1}$d$^{-1}$ in $E_1$ at full reactor $P_{th}$.\\
\indent Collectively, the mentioned background contributions lead to detector-dependent background rates that vary between 6 and 13 counts kg$^{-1}$d$^{-1}$ in the sub-keV$_{ee}$ region $E_1$. Overall background variations are either negligible or can be basically corrected within the validated MC framework.

\section{Conclusions}
\label{chapter7}

The present work provides a comprehensive description of the four large-size sub-keV sensitive HPGe detectors used in the \textsc{Conus} experiment.\\
\indent As demonstrated, it was possible to fulfill all prerequisites needed for observing CE$\nu$NS at a reactor site, assuming realistic quenching factors of nuclear recoils in Ge. The \textsc{Conus} detectors were equipped with low vibration electrical cryocoolers, while large HPGe detector masses of 1\,kg were combined with ultra low noise and very low intrinsic background levels. The achieved pulser resolutions in terms of FWHM lie in the range of [60,~80]\,eV$_{ee}$. These excellent values allow for noise thresholds of the order of 300\,eV$_{ee}$ and full trigger efficiencies down to approximately 200\,eV$_{ee}$. The background levels registered from the \textsc{Conus} detectors inside the \textsc{Conus} shield are around 10 counts kg$^{-1}$d$^{-1}$ in the energy window [0.4,~1.0]\,keV$_{ee}$, which includes a large fraction of the ROI used for CE$\nu$NS searches.\\
\indent Finally, long-term studies revealed a very stable electronic performance over time periods of several months. The noise and background rate stabilities were also investigated. This led to dedicated noise cuts and the possibility of MC-based corrections of the background rate. In this way, stable data periods suitable for CE$\nu$NS detection were established.\\
\indent Optimization work on the \textsc{Conus} setup and its environmental conditions, on the data collection system and data analysis procedures for the last two years of the experiment at the KBR nuclear power plant is progressing. This will further improve the sensitivity of the \textsc{Conus} detectors, which are looking for rare neutrino interactions within and beyond the standard model of elementary particle physics.

\section{Acknowledgements}
\label{sec:acknowlegments}

For technical, mechanical, DAQ, IT and logistical support we thank all involved divisions and workshops at the Max-Planck-Institut f{\"u}r Kernphysik in Heidelberg and at Mirion Technologies (Canberra) in Lingolsheim. We express our deepest gratitude to the Preussen Elektra GmbH for hosting the \textsc{Conus} experiment and for enduring support. We thank J. Westermann for the radon emanation measurement, and Dr. M. Laubenstein for radiopurity assessment of samples used in the \textsc{Conus} detector construction. The \textsc{Conus} experiment is supported financially by the Max Planck Society (MPG). T. Rink is supported by the German Research Foundation (DFG) through the research training group GRK 1940 of the Ruprecht-Karls-University of Heidelberg, and together with J. Hakenm{\"u}ller by the IMPRS-PTFS.

\bibliographystyle{unsrt}

\end{document}